\documentclass[12pt]{article}
\usepackage[dvips]{graphicx}
\usepackage{amsmath,amssymb,amsthm}
\usepackage{setspace}
\usepackage{hyperref}
\usepackage{subfig}
\usepackage{natbib}
\usepackage{color}
\usepackage{array}
\usepackage{float}
\usepackage{endfloat}
\usepackage{ulem}
\usepackage{xcolor}
\DeclareRobustCommand{\erase}{\bgroup\markoverwith{\textcolor{red}{\rule[.5ex]{2pt}{0.4pt}}}\ULon}

\begin{document}


\title{\bf Synchronization analysis between exchange rates on the basis of purchasing power parity using the Hilbert transform}

\author{Makoto Muto
\thanks{Kanda-Misakicho 1--3--2, Chiyoda-ku, Tokyo 101--8360, Japan.
\textit{E-mail}: muto.makoto@nihon-u.ac.jp.}\\
{\scriptsize College of Economics, Nihon University}\\
\and Yoshitaka Saiki
\thanks{Corresponding author. Naka 2--1, Kunitachi-shi, Tokyo 186--8601, Japan. \textit{E-mail}: yoshi.saiki@r.hit-u.ac.jp}\\
{\scriptsize Graduate School of Business Administration, Hitotsubashi University}
}

\date{}

\maketitle


\newpage
\begin{abstract}

\noindent
Synchronization is a phenomenon in which a pair of fluctuations adjust their rhythms when interacting with each other. We measure the degree of synchronization between the U.S. dollar (USD) and euro exchange rates and between the USD and Japanese yen exchange rates on the basis of purchasing power parity (PPP) over time. We employ a method of synchronization analysis using the Hilbert transform, which is common in the field of nonlinear science. We find that the degree of synchronization is high most of the time, suggesting the establishment of PPP. The degree of synchronization does not remain high across periods with economic events with asymmetric effects, such as the U.S. real estate bubble.

\vspace{4ex}

\vspace{2ex}

\end{abstract}

\newpage
\section{Introduction}
\label{Introduction}

Exchange rates are attributed to various theories, such as purchasing power parity (PPP) (\citealp{Cassel_1918}) and interest rate parity (IRP) (\citealp{Keynes_1923}).\footnote{
In addition to PPP and IRP, the flexible-price monetary model (\citealp{Frenkel_1976}) and sticky-price monetary model (\citealp{Dornbusch_1976}) are used. 
Appendix A provides additional information on the theory of exchange rate determination.
}
These theories hold for different time scales; PPP is a long-term theory, whereas IRP is a short-term theory. If PPP holds, then the exchange rate adjusts a deviation from the PPP level within several years. (see \citealp{Rogoff_1996}). \cite{Ito_1997} and \cite{Taylor_2002} apply a unit root test to real exchange rates. \cite{Hasan_2006} applies a unit root test and a cointegration test to the real exchange rate of Australia and Canada. Each study shows that the deviation tends to take several years to adjust to the PPP level depending on the data. \cite{Enders_1994}, \cite{Sarno_1997}, \cite{Ogawa_2008}, and \cite{Mishra_2010} analyze long-term equilibrium values of PPP for multiple real exchange rates.

However, the degree to which the exchange rate theory holds may differ between periods. In other words, the exchange rate adjusts to the equilibrium value indicated by the theory in a certain period but dose not during other periods. Thus, this study quantifies the degree to which the exchange rate theory holds in each period. We focus on exchange rate synchronization on the basis of PPP.\footnote{
IRP analysis requires high-frequency tick data and will be a future task due to data constraints.
}
Synchronization is a phenomenon in which a pair of fluctuations adjust their rhythms when interacting with each other; that is, the phase difference between the two fluctuations remains constant during a certain time interval. Approximately, phase denotes a specific position $(-\pi, \pi]$ in one amplitude of a large fluctuation. The synchronization analysis quantifies the degree of rhythm adjustment between two time series. The following studies apply the synchronization concept to economic analysis. \cite{Flood_2010}, \cite{Ikeda_2013}, and \cite{Esashi_2018} apply synchronization to study business cycles (see \citealp{Onozaki_2018}). \cite{Vodenska_2016} perform a synchronization analysis to examine the interaction and lead-lag relationship between the stock market and the foreign exchange market. \cite{Walti_2011} investigates the relationship between stock market co-movements and monetary integration.

When two numeraire-denominated exchange rates synchronize around the PPP level, we consider that the ratio of the two data adjusts to the original exchange rate’s PPP level. Therefore, the degree of synchronization measures the impact of PPP on the original exchange rate. The impact of PPP denotes the strength at which the exchange rate adjusts to the PPP level. If factors other than PPP significantly impact exchange rates, then the impact of PPP can be small.

The U.S. dollar (USD) and euro (EUR) and the USD and Japanese yen (JPY) exchange rate are analyzed. A monetary authority’s foreign exchange intervention hinders theory establishment, and mutual influence is necessary in this analysis. The USD, EUR, and JPY are the three most traded currencies and have floating exchange rate systems.

The linkages between the exchange rate and international monetary policy or the degree of openness of international capital markets are now summarized. Linkages between exchange rates may occur even for currencies of countries with less freedom of capital movement and less degree of openness in their international capital markets. However, such linkages are not the subject of our analysis. For example, we consider a country with a fixed exchange rate system (dollar pegged) in which the degree of freedom of capital movement and the openness of international capital markets are low to maintain the currency system. The regulation of capital transactions hinders the establishment of PPP; however, a strong linkage occurs between the country's currency and the dollar. However, this linkage is the result of government intervention and does not underlie the PPP that is the subject of this analysis. Therefore, the currencies in our analysis should be from countries with less government regulation or exchange rate intervention because such a linkage between exchange rates suggests a theoretical background, such as PPP. The United States, the euro area, and Japan fall into this category because of their freedom of capital movement and open international capital markets.

Our analysis shows that the degree of synchronization is stably high between the USD and EUR and between the USD and JPY during certain periods.\footnote{
More precisely, it is the synchronization between the deviations in the exchange rate from the PPP of numeraire-denominated USD and EUR, or that of the denominated USD and JPY. See section 4.3.
}
This result suggests that the USD/EUR and USD/JPY exchange rates adjust to PPP during the period. During other periods, the degree of synchronization does not maintain a high level for either the USD/EUR or USD/JPY. This result suggests that certain factors other than PPP affect the exchange rate during a period.

The remainder of this paper is structured as follows. Section \ref{Synchronization} defines synchronization, and Section \ref{Data} explains the data used in this analysis. Section \ref{PPP and frequency band} provides calculations of exchange rate deviation from PPP and introduces a frequency-based filter. Section \ref{Synchronization analysis using the Hilbert transform} conducts a synchronization analysis using the Hilbert transform. Section \ref{Results and interpretation} provides the results of the synchronization analysis. Section \ref{Comparison with the correlation coefficient} discusses the difference between a synchronization analysis and the correlation coefficient. Finally, Section \ref{Conclusions} concludes our paper.

\section{Synchronization}
\label{Synchronization}

\begin{figure}
    \begin{center}
        \subfloat{({\bf a}) }{\includegraphics[clip, width=0.42\columnwidth,height=0.28\columnwidth]{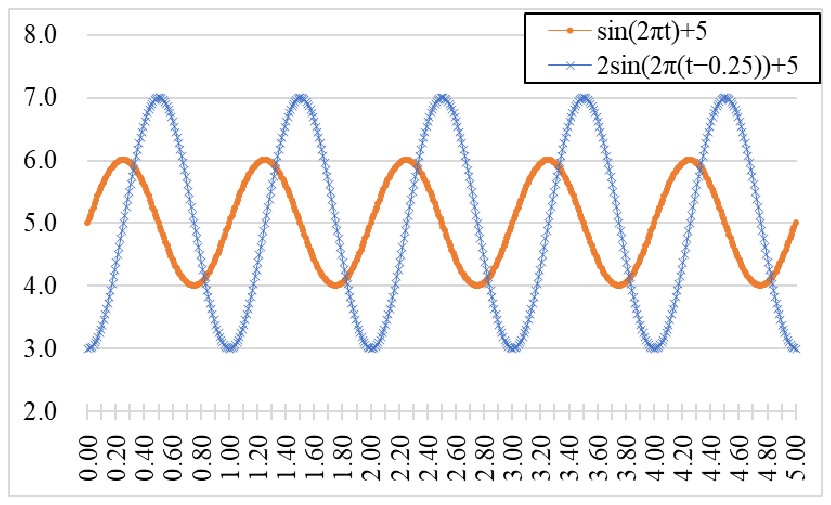}}
        \subfloat{({\bf b}) }{\includegraphics[clip, width=0.42\columnwidth,height=0.28\columnwidth]{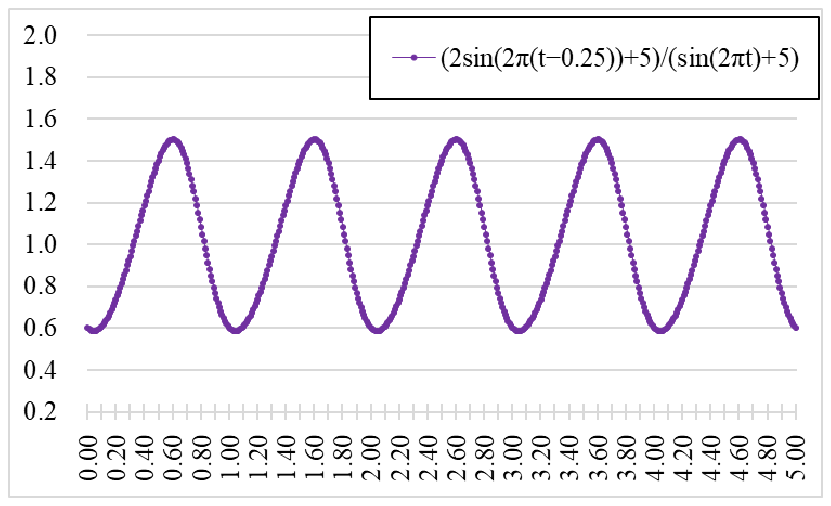}}\\
    \end{center}
    \begin{spacing}{1.1}
        Figure 1. (a) Two phase-synchronized time series. Note: “$\bullet$” (orange) represents $\sin(2\pi t)+5$, and “$\times$” (blue) represents $2\sin(2\pi(t-0.25))+5$. The amplitudes of two time series are different. Although the phases of two time series are also different from each other, the phase difference between two time series is kept constant.
        (b) The ratio of the two synchronized time series. Note: “$\bullet$” (purple) represents $\frac{2\sin(2\pi(t-0.25))+5}{\sin(2\pi t)+5}$. This ratio oscillates around 1.
    \end{spacing}
    \label{fig:f1}
\end{figure}
Synchronization is a phenomenon in which a pair of fluctuations adjust their rhythms through mutual interactions. Rhythm adjustment indicates that the phase difference between two time series adjusts to remain constant for a certain time interval. We use the notion of phase to capture the degree of synchronization. The synchronization concept is explained using the simple vibrations, $f(t)=\sin(2\pi t)+5$ and $g(t)=2\sin(2\pi(t-0.25))+5$ [Figure 1(a)]. Phase represents a specific position $(-\pi,\pi]$ in one amplitude of oscillation data.\footnote{
For details, see Section \ref{Hilbert transform and instantaneous phase} for the definition of the phase.
}
The phases of $f(t)$ and $g(t)$ are $2\pi t$ and $2\pi(t-0.25)$, respectively, in which each phase is converted into $(-\pi,\pi]$ value. Two time series, namely, $f(t)$ and $g(t)$, are synchronized if the phase difference is constant in time.
This synchronization is called phase synchronization. In the previous case, the phase difference is $0.5n$ for all $t$, which are synchronized. In the Figure 1(b) illustrates the ratio of two synchronized time series, whose ratio $\frac{2\sin(2\pi(t-0.25))+5}{\sin(2\pi t)+5}$ fluctuates around a reference value of approximately 1. In other words, this ratio has the property of returning to the reference value. This property is used to quantify the strength of the exchange rate returning to the PPP level.

\section{Data}
\label{Data}

Monthly average nominal exchange rate data for the USD/EUR, USD/JPY, Australian dollar (AUD)/USD, AUD/EUR, AUD/JPY, New Zealand dollar (NZD)/USD, NZD/EUR, and NZD/JPY during 1999:01--2017:12, 1987:01--2017:12, 1999:01--2017:12, 1999:01--2017:12, 1987:01--2017:12, 1999:01\\\noindent--2017:12, 1999:01--2017:12, and 1987:01--2017:12, respectively, are used. These data are obtained from \textit{Datastream}. Monthly producer price index data are used as the price level data of the United States, the euro area, and Japan, for which 2010 data are normalized to 100. These data are obtained from the \textit{International Financial Statistics} of the International Monetary Fund (IMF) website. Current account balances (percent of GDP) of the United States, the euro area, and Japan are yearly data obtained from the \textit{World Economic Outlook} of the IMF website.

\section{PPP and frequency band}
\label{PPP and frequency band}

\subsection{PPP}\label{PPP}
The PPP level is calculated as
\begin{equation}
    \rho_t^{USDj}=S_{base}^{USDj}\frac{P_t^j/P_{base}^j}{P_t^{USD}/P_{base}^{USD}},
    \label{eq:PPP}
\end{equation}
where $S_t^{USDj}$ denotes the USD/currency $j$ ($j=$EUR, JPY) exchange rate at time $t$, $S_{base}^{USDj}$ denotes the USD/currency $j$ ($j=$EUR, JPY) exchange rate at a base time, $P_t^j$ denotes the country of currency $j$'s price index at time $t$, and $P_{base}^j$ denotes the country of currency $j$'s price index at the base time. To calculate PPP, the base time is selected when the current account balances are close to zero, $\rho_t^{USDEUR}$ is in 2010, and $\rho_t^{USDJPY}$ is in 1991. Exchange rate fluctuations around PPP are calculated as
\begin{equation}
     \xi_t^{USDj}:=\frac{S_t^{USDj}}{\rho_t^{USDj}}.
    \label{eq:FPPP}
\end{equation}

Figure 2(a) represents the USD/EUR exchange rate and PPP level. When the USD/EUR exchange rate deviates from the PPP level, it returns to its level in approximately one to five years. Since 2015, the exchange rate has deviated from the PPP level. Figure 2(b) represents the divergence of the USD/EUR exchange rate from PPP, which fluctuates around 1. Similarly, Figure 2(c) represents the USD/JPY exchange rate and PPP level. The USD/JPY exchange rate returns to the PPP level approximately every two to five years. In addition, the return time is longer than that of the USD/EUR exchange rate. Since 2013, the exchange rate has deviated from the PPP level. Figure 2(d) represents the divergence of the USD/JPY exchange rate from the PPP level, which fluctuates around 1.

\begin{figure}
    \begin{center}
        \subfloat{({\bf a}) }{\includegraphics[clip, width=0.435\columnwidth,height=0.31\columnwidth]{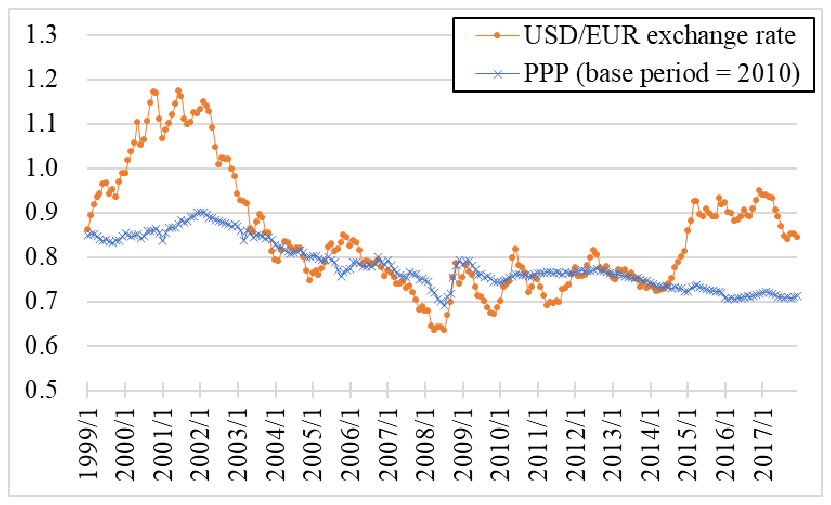}}
        \subfloat{({\bf b}) }{\includegraphics[clip, width=0.435\columnwidth,height=0.31\columnwidth]{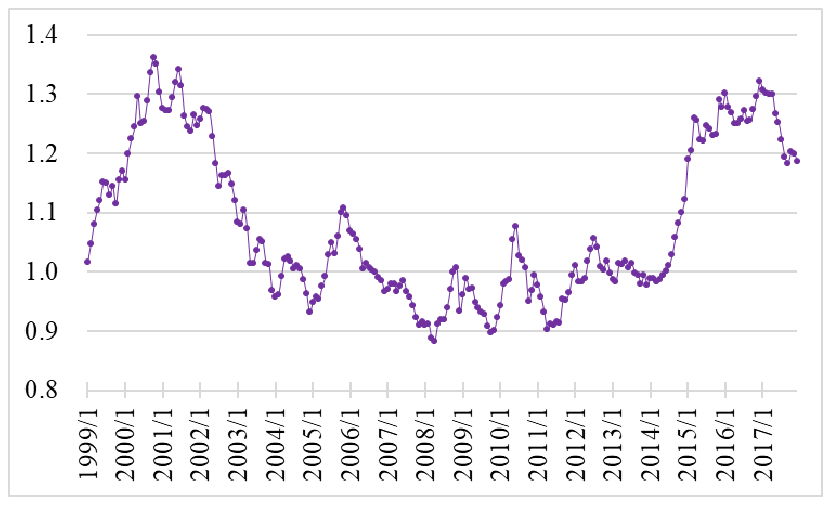}}\\
        \vspace{2mm}
        \subfloat{({\bf c}) }{\includegraphics[clip, width=0.435\columnwidth,height=0.31\columnwidth]{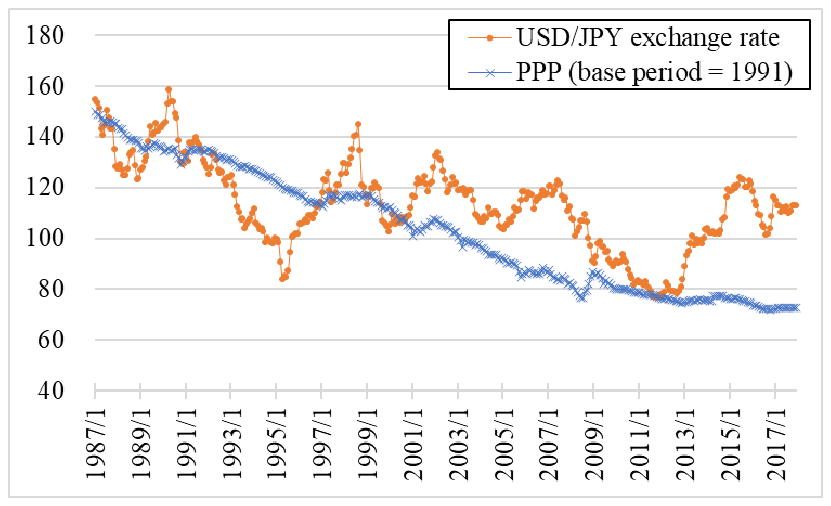}}
        \subfloat{({\bf d}) }{\includegraphics[clip, width=0.435\columnwidth,height=0.31\columnwidth]{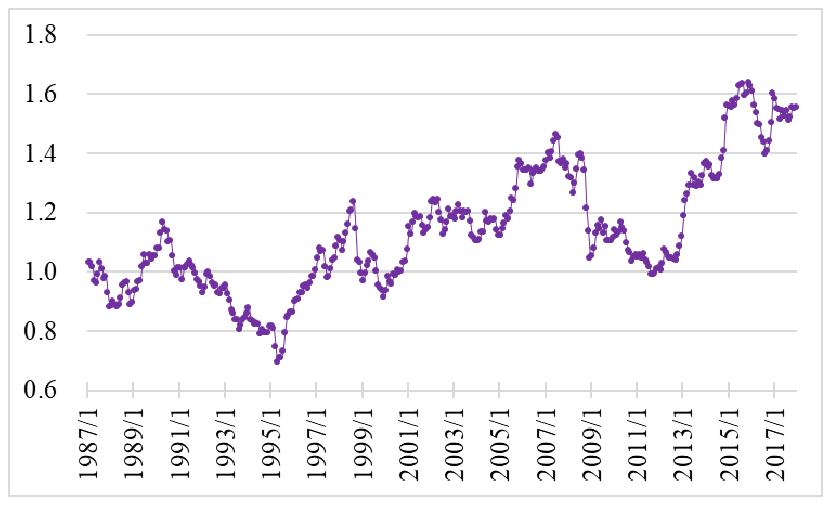}}
    \end{center}
    \begin{spacing}{1.1}
        Figure 2. (a) USD/EUR exchange rate and PPP level ($\rho_t^{USDEUR}$). Note: “$\bullet$” (orange) represents the USD/EUR exchange rate, and “$\times$” (blue) represents the PPP level. When the USD/EUR exchange rate deviates from the PPP level, it returns to its level in approximately one to five years.
        (b) The exchange rate fluctuations around PPP ($\xi_t^{USDEUR}$).
        Note: “$\bullet$" (purple) represents the divergence of the USD/EUR exchange rate from PPP. The value fluctuates around 1.
        (c) USD/JPY exchange rate and PPP level ($\rho_t^{USDJPY}$). Note: “$\bullet$” (orange) represents the USD/JPY exchange rate, and “$\times$” (blue) represents the PPP level. The USD/JPY exchange rate has returned to the PPP level over a long period.The USD/JPY exchange rate returns to the PPP level approximately every two to five years.
        (d) The exchange rate fluctuations around PPP ($\xi_t^{USDJPY}$). Note: “$\bullet$” (purple) represents the divergence of the USD/JPY exchange rate from the PPP level. The value fluctuates around 1.
    \end{spacing}
    \label{fig:f2}
\end{figure}

\subsection{Power spectrum}
\label{Power spectrum}

Exchange rates have many determinants other than PPP. If data contain fluctuations of various sizes, then the appropriate phase cannot be clearly defined. Therefore, identifying the frequency band that holds PPP and generates data with the frequency is needed. The power spectrum is used to identify the frequency band of the exchange rate fluctuation around PPP.
See Appendix B Eq.~(\ref{eq:A9}) for the definition of the power spectrum.
Figure 3(a) shows the log power of $ \xi_t^{USDEUR}$, where the horizontal axis represents the monthly frequency. For example, the leftmost part of the figure suggests that a frequency component with amplitude 228 months has the highest power. The arrow represents the band used in this analysis. Figure 3(b) shows the log power of $ \xi_t^{USDJPY}$.

The power spectrum implies that the USD/EUR exchange rate fluctuates around PPP in frequencies ranges of approximately 32.6--228.0 and 38.0--228.0 months. Then, the time series is extracted with the frequency band of that range using the lower cutoff frequency of $k_0=1$ and upper cutoff frequency of $k_1=7$, corresponding to 228.0 $(\approx228/k_0)$ and 32.6 $(\approx228/k_1)$ months, respectively. (See Appendix B for the definition of $k_0$ and $k_1$.) Thus, if PPP holds, then deviations of the USD/EUR exchange rate from PPP have vanished from 16.3 (32.6/2) to 114.0 (228.0/2) months. Therefore, in the USD/EUR analisys, we focus on the range of the frequency band of 32.6--228.0 and 38.0--228.0 months.
Similarly, the corresponding frequency bands of 37.2--186.0 or 41.3--186.0 months are expected for the USD/JPY exchange rate. Thus, if PPP holds, then the USD/JPY exchange rate deviations from PPP vanish from 18.6 (37.2/2) to 93.0 (186.0/2) months. We focus on 37.2--186.0 and 41.3--86.0 months as the frequency band in the USD and JPY analysis. The band movements are considered to represent price adjustments through trade and productivity adjustments.

\begin{figure}
    \begin{center}
        \subfloat{({\bf a}) }{\includegraphics[clip, width=0.435\columnwidth,height=0.31\columnwidth]{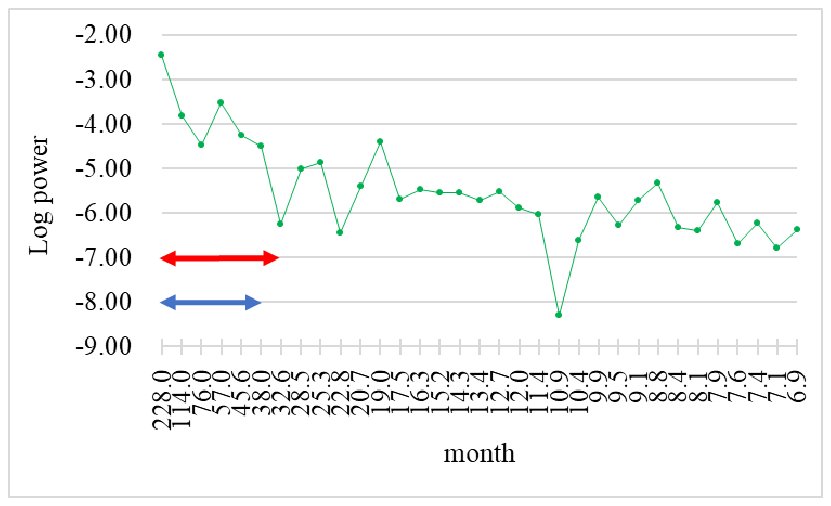}}
        \subfloat{({\bf b}) }{\includegraphics[clip, width=0.435\columnwidth,height=0.31\columnwidth]{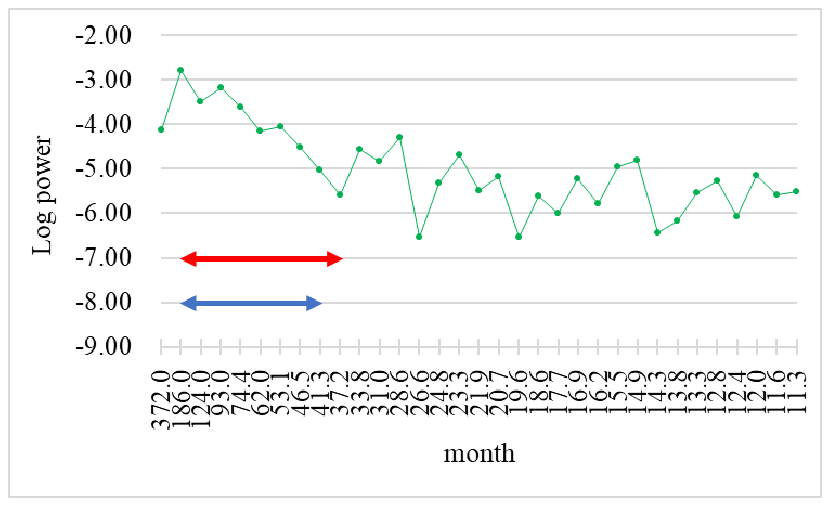}}\\
    \end{center}
    \begin{spacing}{1.1}
        Figure 3. (a) The log power of the exchange rate fluctuations around PPP ($\xi_t^{USDEUR}$). Note: The horizontal axis represents monthly frequency. The arrow represents the band used in this analysis. The exchange rate fluctuations around PPP ($\xi_t^{USDEUR}$) oscillate in the frequency band represented by the arrows. Therefore, we expect that the USD/EUR exchange rate fluctuates around PPP in approximate frequency ranges of 32.6--228.0 and 38.0--228.0.
        (b) The log power of the exchange rate fluctuations around PPP ($\xi_t^{USDJPY}$). Note: We can expect that the USD/EUR exchange rate fluctuates around PPP ($\xi_t^{USDJPY}$) in the approximate frequency ranges of 37.2--186.0 or 41.3--186.0.
    \end{spacing}
    \label{fig:f3}
\end{figure}

\subsection{Numeraire}
\label{Numeraire}

We analyze the synchronization between fluctuations in the USD and currency $j$ of $\xi_t^{USDj}$ using the numeraire to divide $\xi_t^{USDj}$ into USD and currency $j$ parts. Then, the synchronization between these two time series is studied. \cite{Frankel_1994} and \cite{McKinnon_2004} adopted this method to analyze the linkage between multiple exchange rates using numeraire-denominated exchange rates.

$\xi_t^{USDj}$ can be written using the AUD numeraire as\footnote{
Numeraire often uses the currency of the floating exchange rate system. The Swiss franc (CHF) and NZD are also often used for numeraire. We do not use the CHF because Swiss National Bank sets minimum exchange rate at CHF 1.20 per EUR from 2011 to 2015. Changing the NZD to numeraire does not affect the result. See Appendix C for details.
}
\begin{equation}
     \xi_t^{USDj}\approx\frac{\xi_t^{AUDj}}{\xi_t^{AUDUSD}}=\frac{\frac{S_t^{AUDj}}{S_{base}^{AUDj}({P_t^j}/{P_{base}^j})}}{\frac{S_t^{AUDUSD}}{S_{base}^{AUDUSD}({P_t^{USD}}/{P_{base}^{USD}})}}=\frac{\xi^{\prime~AUDj}_t}{\xi^{\prime~AUDUSD}_t},
    \label{eq:FPPP2}
\end{equation}
where 
$$\xi^{\prime~AUDUSD}_t:=\frac{S_t^{AUDUSD}}{S_{base}^{AUDUSD}({P_t^{USD}}/{P_{base}^{USD}})},$$ 
and
$$\xi^{\prime~AUDj}_t:=\frac{S_t^{AUDj}}{S_{base}^{AUDj}({P_t^j}/{P_{base}^j})}.$$ $\xi^{\prime~AUDUSD}_t$
\footnote{$\xi^\prime_t$ is not PPP in a strict sense because it excludes the inflation rate of numeraire. However, we use $ \xi^\prime_t$ because of the small sample size of the PPI data of Australia. As shown in 
Eq.~\eqref{eq:FPPP2}, this condition does not affect the analysis because the $\xi^\prime_t$ ratio is the same as that of $ \xi_t$.
}
and $\xi^{\prime~AUDj}_t$ exclude the AUD inflation rate from $ \xi_t^{AUDUSD}$ and $\xi_t^{AUDj}$. When $\xi^{\prime~AUDUSD}_t$ and $\xi^{\prime~AUDj}_t$ have the same fluctuation, $\xi_t^{USDj}$ fluctuates around 1 (see Figure 1), thus suggesting the establishment of PPP.

\subsection{Band-pass filter}
\label{Band-pass filter}

We extract a frequency band of a numeraire-denominated exchange rate using PPP as previously described. Time series data are generated with a frequency estimated from each $\xi^{\prime~AUDUSD}_t$ and $\xi^{\prime~AUDj}_t$ using a band-pass filter.\footnote{
See Appendix B for the band-pass filter details. \cite{Rodriguez_1999} and \cite{Varela_2001} perform a synchronization analysis using band-pass-filtered data for brain science research. Business cycle studies often use band-pass-filtered economic time series data \citep{Baxter_1999, Calderon_2007}.
}
Figure 4(a) shows $ \eta^{\prime~AUDUSD}_t$ and $\eta^{\prime~AUDEUR}_t$ with a 32.6--228.0-month band-pass filter applied to $\xi^{\prime~AUDUSD}_t$ and $\xi^{\prime~AUDEUR}_t$. 
The two time series fluctuation rhythms do not adjust around year 2015 but in other periods. 
The two time series fluctuation rhythms do not adjust around 2015 but do so during other periods. Figure 4(b) indicates $ \eta^{\prime~AUDUSD}_t$ and $\eta^{\prime~AUDEUR}_t$ with a 38.0--228.0-month band-pass filter applied to $\xi^{\prime~AUDUSD}_t$ and $\xi^{\prime~AUDEUR}_t$. The fluctuation
rhythms of the two time series do not adjust around 2006 but do so during other periods.
Figure 4(c) shows $\eta^{\prime~AUDUSD}_t$ and $\eta^{\prime~AUDJPY}_t$ with a 37.2--186.0-month band-pass filter applied to $\xi^{\prime~AUDUSD}_t$ and $\xi^{\prime~AUDJPY}_t$. The two time series fluctuation rhythms adjust around 2001 but do not do so during other periods. Figure 4(d) shows $\eta^{\prime~AUDUSD}_t$ and $\eta^{\prime~AUDJPY}_t$ with a 41.3--186.0-month band-pass filter applied to $\xi^{\prime~AUDUSD}_t$ and $\xi^{\prime~AUDJPY}_t$. The two time series fluctuation rhythms adjust around 2001 but do not do so during other periods.
\begin{figure}
    \begin{center}
        \subfloat{({\bf a}) }{\includegraphics[clip, width=0.42\columnwidth,height=0.28\columnwidth]{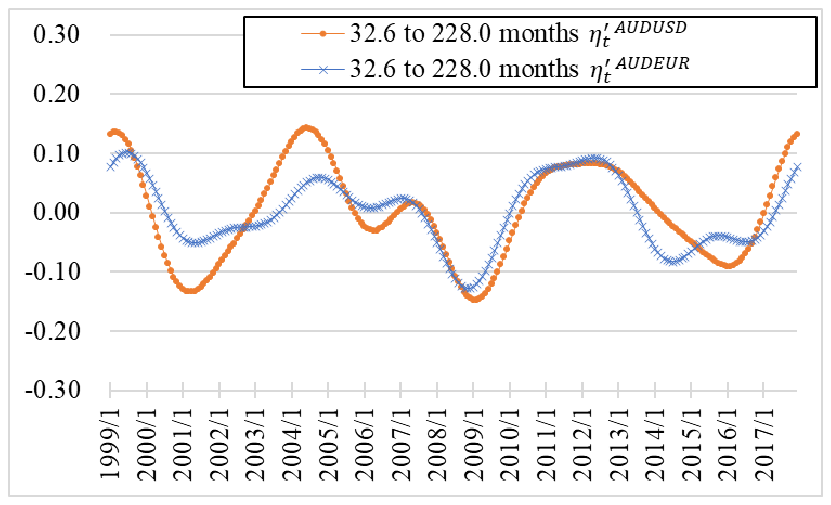}}
        \subfloat{({\bf b}) }{\includegraphics[clip, width=0.42\columnwidth,height=0.28\columnwidth]{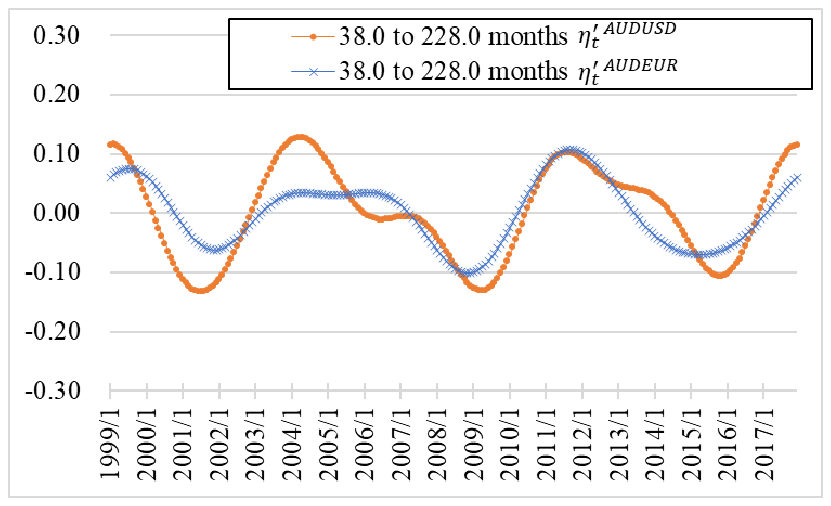}}\\
        \vspace{2mm}
        \subfloat{({\bf c}) }{\includegraphics[clip, width=0.42\columnwidth,height=0.28\columnwidth]{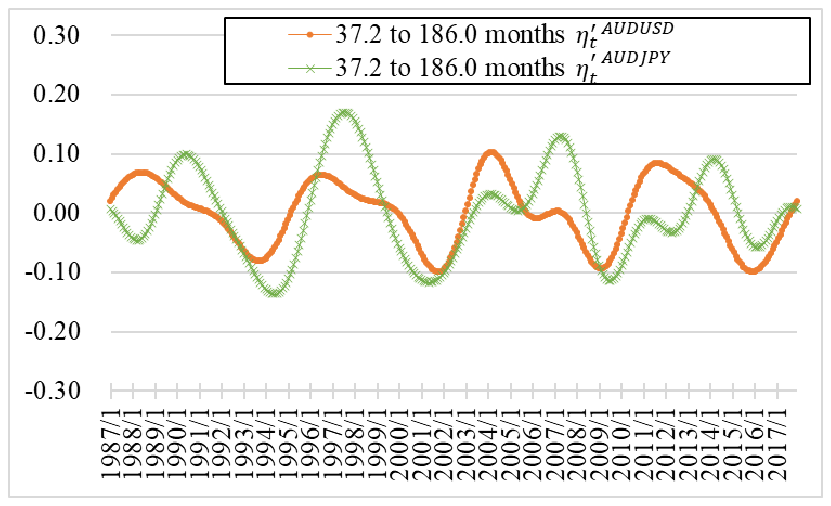}}
        \subfloat{({\bf d}) }{\includegraphics[clip, width=0.42\columnwidth,height=0.28\columnwidth]{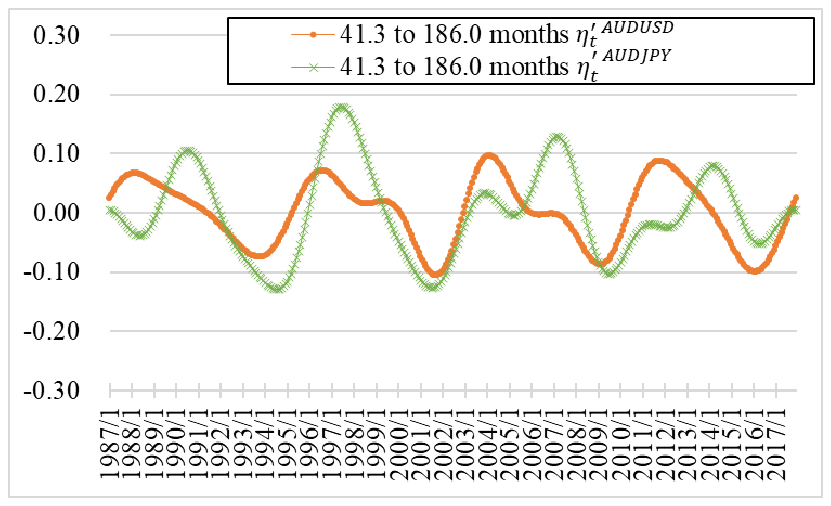}}
    \end{center}
    \begin{spacing}{1.1}
        Figure 4. (a) The 32.6--228.0-month band-pass filtered data $\eta^{\prime~AUDUSD}_t$ and $\eta^{\prime~AUDEUR}_t$ of the exchange rate fluctuations around PPP ($\xi^{\prime~AUDUSD}_t$ and $\xi^{\prime~AUDEUR}_t$). Note: The two time series fluctuation rhythms do not adjust during year 2015 but during other periods.
        (b) The 38.0--228.0-month band-pass filtered data $\eta^{\prime~AUDUSD}_t$ and $\eta^{\prime~AUDEUR}_t$ of the exchange rate fluctuations around PPP ($\xi^{\prime~AUDUSD}_t$ and $\xi^{\prime~AUDEUR}_t$). Note: The two time series fluctuation rhythms do not adjust during 2006 but during other periods.
        (c) The 37.2--186.0-month band-pass filtered data $\eta^{\prime~AUDUSD}_t$ and $\eta^{\prime~AUDJPY}_t$ of the exchange rate fluctuations around PPP ($\xi^{\prime~AUDUSD}_t$ and $\xi^{\prime~AUDJPY}_t$). Note: The two time series fluctuation rhythms adjust during 2001 but not during other periods.
        (d) The 41.3--186.0-month band-pass filtered data $\eta^{\prime~AUDUSD}_t$ and $\eta^{\prime~AUDJPY}_t$ of the exchange rate fluctuations around PPP ($\xi^{\prime~AUDUSD}_t$ and $\xi^{\prime~AUDJPY}_t$). Note: The two time series fluctuation rhythms adjust during 2001 but not during other periods.
    \end{spacing}
    \label{fig:f4}
\end{figure}

\section{Synchronization analysis using the Hilbert transform}
\label{Synchronization analysis using the Hilbert transform}

\subsection{Hilbert transform and instantaneous phase}
\label{Hilbert transform and instantaneous phase}

In this paper, the Hilbert transform is used in our synchronization analysis to define a phase at each time. Some previous studies applied the Hilbert transform to an analysis of macroeconomic data. \cite{Ikeda_2013} analyze the business cycle in Japan using Indices of Industrial Production (IIP) data for 16 industrial sectors. They use the Hilbert transform to calculate phases from the data and find a partial phase synchronization of Japan's business cycles. \cite{Vodenska_2016} study the interactions between global equity and foreign exchange markets by analyzing daily price data for foreign exchange markets and major stock indices for 48 countries. They use the complex Hilbert principal component analysis (CHPCA), which is a complex version of principal component analysis that uses Hilbert transformed values for the imaginary part of the complex time series. They show that the information on lead-lag relationships in financial markets and exchange rates obtained by CHPCA can serve as an early warning system (EWS) against systemic risk contagion. The Hilbert transform is also applied to analyses in the finance field. \cite{Fusai_2016} proposed the Wiener-Hopf factorization of complex functions that can be applied to the pricing of barrier and look-back options when monitoring is discrete. This method uses the Hilbert and z-transforms. \cite{Phelan_2018} modify the pricing of options pricing using the discrete monitoring proposed by \cite{Fusai_2016} when monitoring is continuous. In addition, they examine the truncation error of the sinc-based Hilbert transform to investigate the error of the pricing method. \cite{Phelan_2019} improve the option pricing method using the sinc-based Hilbert transform with a spectral filter. This method can be applied to, for example, the pricing of barrier options when monitoring is discrete. \cite{Phelan_2020} presents new pricing methods for Bermuda, American, and $\alpha$-quantile options that use the Hilbert transforms, which have the advantage of small errors and fast CPU time.

We define a phase to measure the phase difference and assume that $\eta^{\prime~AUDUSD}_t$ and $\eta^{\prime~AUDj}_t$ are obtained from the real part of the complex variable data. The imaginary part of the complex data can be generated using the Hilbert transform value of the real part. The Hilbert transform can be realized by an ideal ﬁlter, for which the amplitude and phase responses are unity and a constant $\pi/2$ lag at all Fourier frequencies (\cite{Pikovsky_2001}, pp. 362-363). The Hilbert transform is expressed as 
\begin{equation}
     s_t^H=\frac{1}{\pi}P.V.\int_{-\infty}^{\infty}{\frac{s_\tau}{t-\tau}d\tau,}
    \label{eq:Hilbert}
\end{equation}
where $s_t$ denotes the time series data at time $t$ and $P.V.$ denotes the Cauchy principal value integrals.
A complex valued time series ${\hat{s}}_t$ is constructed whose real part is actual data $s_t$, and the imaginary part $s_t^H$ is generated from $s_t$ using the Hilbert transform:
\begin{equation}
    {\hat{s}}_t=s_t+s_t^Hi.
    \label{eq:complex}
\end{equation}
The frequency spectrum of the complex number is as follows:
\begin{equation}
    \hat{S}\left(f\right)=
    \left\{\begin{matrix}2S(f)\\S(f)\\0\\\end{matrix}\begin{matrix}\\\\\\\end{matrix}\right.\begin{matrix}f>0\\f=0\\f<0\\\end{matrix}
    \label{eq:complex_f}
\end{equation}
where $\hat{S}\left(f\right)$ is the Fourier transform of ${\hat{s}}_t$, and $S\left(f\right)$ is the Fourier transform of $s_t$. We compute the ${\hat{s}}_t$ using forward and inverse Fourier transforms. Specifically, the formula is as follows:
\begin{equation}
    {\hat{s}}_t=F^{-1}\left(S\left(f\right)2U\right)=s_t+s_t^Hi
    \label{eq:complex_c}
\end{equation}
where $F^{-1}$ is an inverse Fourier transform, and $U$ is the unit step function.\footnote{
We use the python program “scipy.signal.Hilbert” to obtain the Hilbert transform value from this calculation. This approach can be further refined with a sinc functions expansion yielding exponential rather than polynomial convergence of the error on the grid size, as explained in \cite{Fusai_2016}.
}
This calculation provides the Hilbert transform value $s_t^H$.

Phase is defined by the angle $\phi_t$ formed by the horizontal axis and complex variable data
\begin{equation}
    \phi_t=  
    \begin{cases}
        \tan^{-1}\left(\frac{s_t^H}{s_t}\right) & (st>0)\\
        \tan^{-1}\left(\frac{s_t^H}{s_t}\right)+\pi & (st<0)
    \end{cases}
    .
    \label{eq:Phase}
\end{equation}
Phase at a certain time is called an instantaneous phase. The value can be discontinuous over time because it ranges from $-\pi$ to $\pi$. Figure 5(a) shows that $s_t=\sin(2\pi t)$, and its Hilbert transform value $s_t^H=\sin(2\pi(t-0.25))$, which implies $s_t^H=s_{t-\frac{\pi}{2}}$. Figure 5(b) shows a behavior of $(s_t, s_t^H)$ in a complex plane.\footnote{An outlier occurs at both ends of the Hilbert transform values when numerical computations are performed. Thus, certain complex variable 
data near $(0,-1)$ in the complex plane deviate slightly from the 
unit circle. Therefore, we exclude 10 data points from both ends in the following analysis.
}
Phase is identified by the angle $\phi_t$ formed by the real axis and complex variable data.

\begin{figure}
    \begin{center}
        \subfloat{({\bf a}) }{\includegraphics[clip, width=0.42\columnwidth,height=0.28\columnwidth]{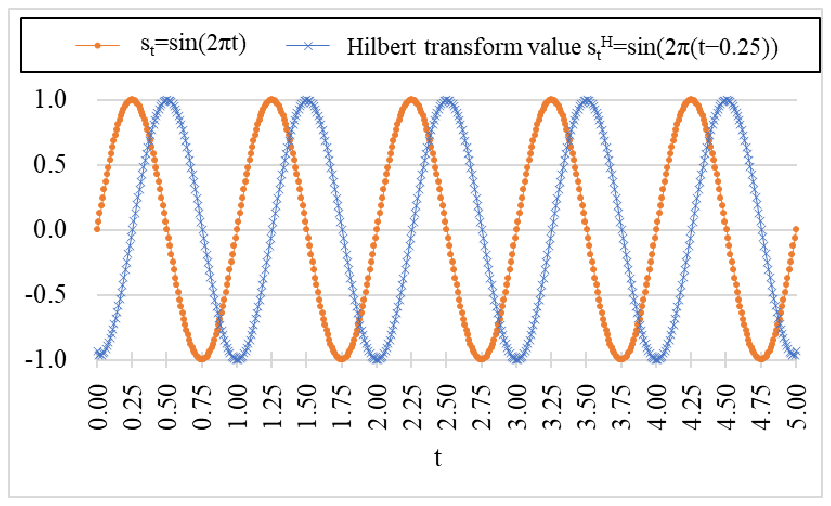}}
        \subfloat{({\bf b}) }{\includegraphics[clip, width=0.28\columnwidth,height=0.28\columnwidth]{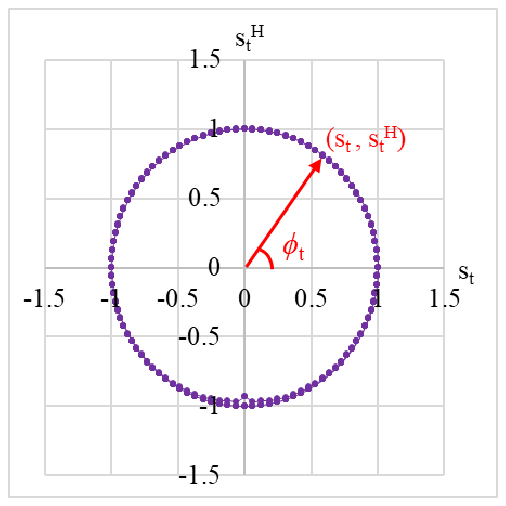}}\\
    \end{center}
    \begin{spacing}{1.1}
        Figure 5. (a) $s_t=\sin(2\pi t)$ and its Hilbert transform value   $s_t^H=\sin(2\pi(t-0.25))$. Note: “$\times$” (blue) represents $s_t^H=\sin(2\pi(t-0.25))$ delayed by $\pi/2$ from $s_t=\sin(2\pi t)$ and is created by the Hilbert transform. (b) Behavior of $(s_t,\  s_t^H)$ on a complex plane. Note: The phase is defined by the angle $\phi_t$ formed by the horizontal axis and the complex variable data $(s_t,\  s_t^H)$.
    \end{spacing}
    \label{fig:f5}
\end{figure}

Figures 6(a) and (b) show the behavior of $(s_t, s_t^H)$, where $s_t=\eta^{\prime~AUDUSD}_t$ and $\eta^{\prime~AUDEUR}_t$ with 32.6--228.0 months, respectively. Figures 6(c) and (d) show the behavior of $(s_t, s_t^H)$, where $s_t=\eta^{\prime~AUDUSD}_t$ and  $\eta^{\prime~AUDJPY}_t$ with 37.2--186.0 months, respectively.

\begin{figure}
    \begin{center}
        \subfloat{({\bf a}) }{\includegraphics[clip, width=0.28\columnwidth,height=0.28\columnwidth]{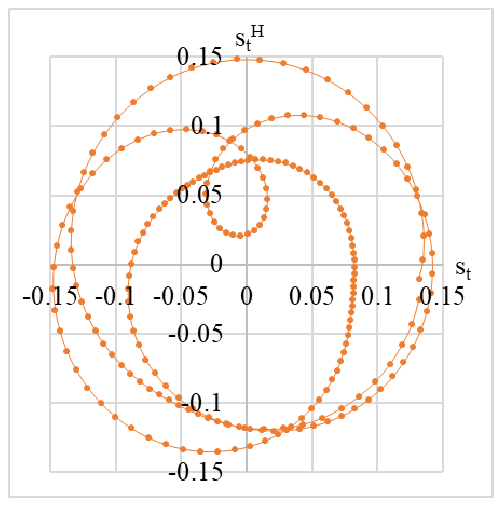}}
        \subfloat{({\bf b}) }{\includegraphics[clip, width=0.28\columnwidth,height=0.28\columnwidth]{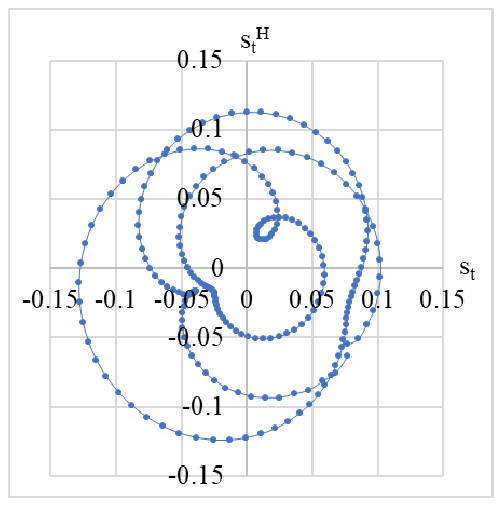}}\\
        \vspace{2mm}
        \subfloat{({\bf c}) }{\includegraphics[clip, width=0.28\columnwidth,height=0.28\columnwidth]{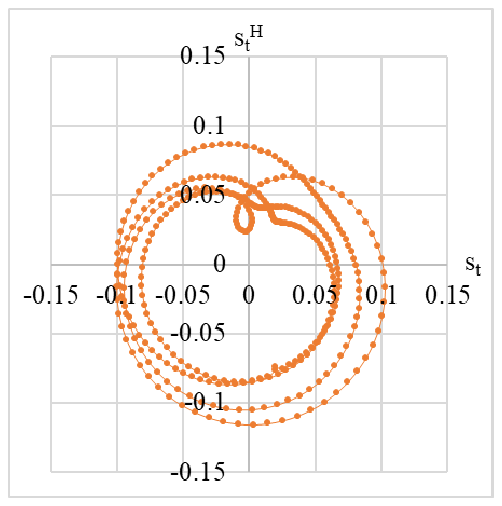}}
        \subfloat{({\bf d}) }{\includegraphics[clip, width=0.28\columnwidth,height=0.28\columnwidth]{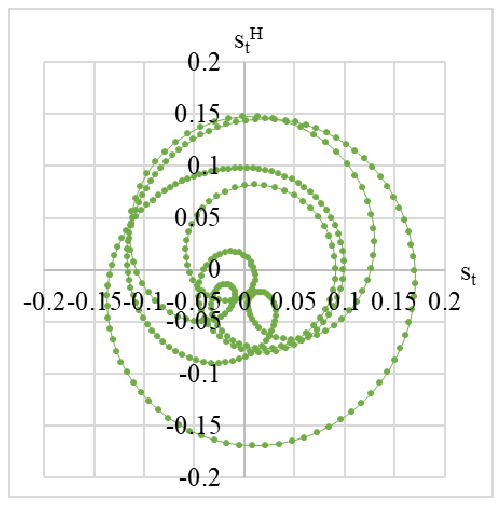}}
    \end{center}
    \begin{spacing}{1.1}
        Figure 6. (a) Behavior of $(s_t,\  s_t^H)$, where $s_t=\eta^{\prime~AUDUSD}_t$ with 32.6--228.0 months. (b) Behavior of $(s_t,\ s_t^H)$, where $s_t=\eta^{\prime~AUDEUR}_t$ with 32.6--228.0 months. (c) Behavior of $(s_t,\ s_t^H)$, where   $s_t=\eta^{\prime~AUDUSD}_t$ with 37.2--186.0 months. (d) Behavior of $(s_t,\ s_t^H)$, where $s_t=\eta^{\prime~AUDJPY}_t$ with 37.2--186.0 months. Note: When defining the phase, the complex data should rotate around the origin. The phase can jump when the complex data move quite close to the origin. From the figure, the complex data seem to rotate around the origin, although a few small circles are mixed in.
        \end{spacing}
    \label{fig:f6}
\end{figure}

Figures 7(a)--(d) show instantaneous phases of 32.6--228.0, 38.0--228.0, 37.2--186.0, and 41.3--186.0 months for $\eta^{\prime~AUDUSD}_t$ and $\eta^{\prime~AUDEUR}_t$, $\eta^{\prime~AUDUSD}_t$ and $\eta^{\prime~AUDEUR}_t$, $\eta^{\prime~AUDUSD}_t$ and $\eta^{\prime~AUDJPY}_t$, and $\eta^{\prime~AUDUSD}_t$ and $\eta^{\prime~AUDJPY}_t$, respectively. The phase differences in Figures 7(a)--(d) are not constant around 2006 (a), 2006 (b), 2005 and 2012 (c), and 2005 (d), respectively; however, all of them are almost constant during in other periods.

\begin{figure}
    \begin{center}
        \subfloat{({\bf a}) }{\includegraphics[clip, width=0.42\columnwidth,height=0.28\columnwidth]{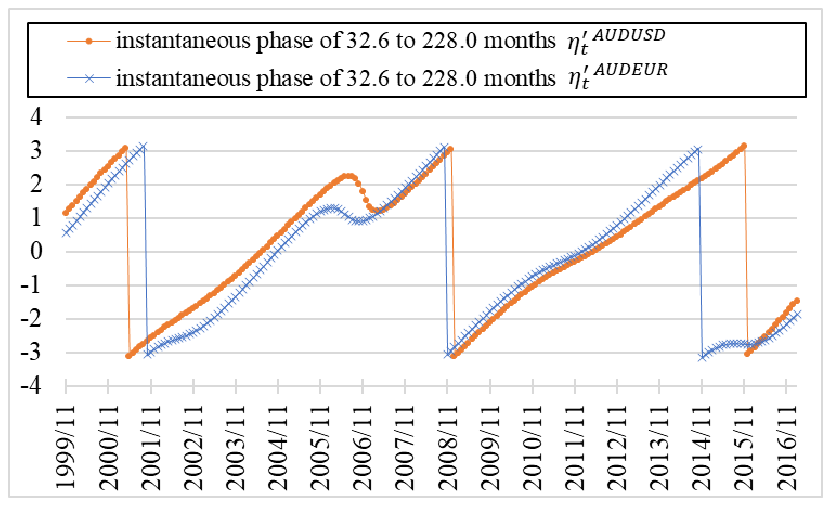}}
        \subfloat{({\bf b}) }{\includegraphics[clip, width=0.42\columnwidth,height=0.28\columnwidth]{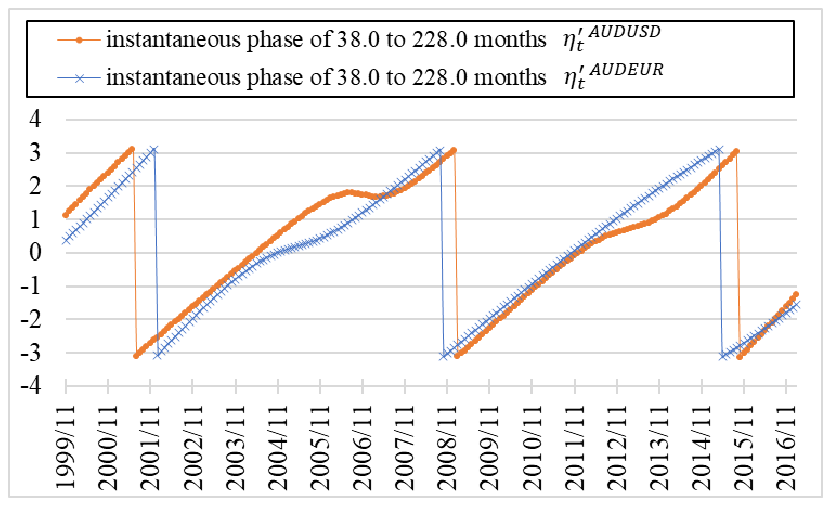}}\\
        \vspace{2mm}
        \subfloat{({\bf c}) }{\includegraphics[clip, width=0.42\columnwidth,height=0.28\columnwidth]{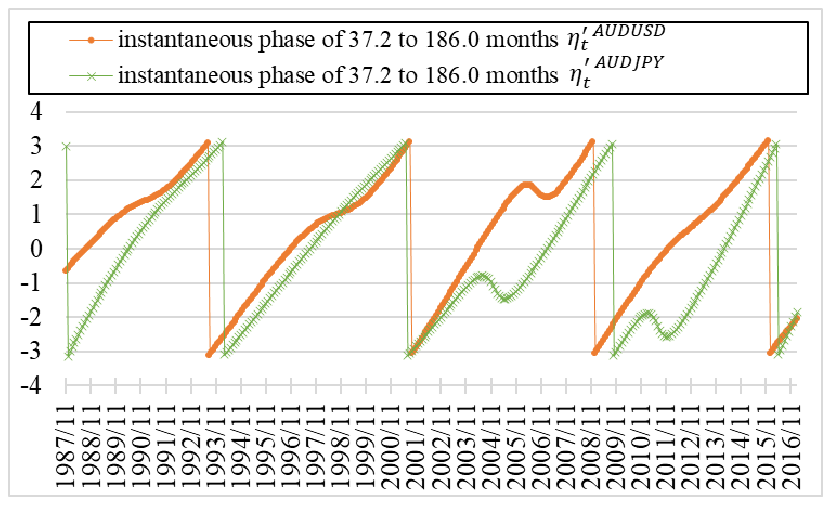}}
        \subfloat{({\bf d}) }{\includegraphics[clip, width=0.42\columnwidth,height=0.28\columnwidth]{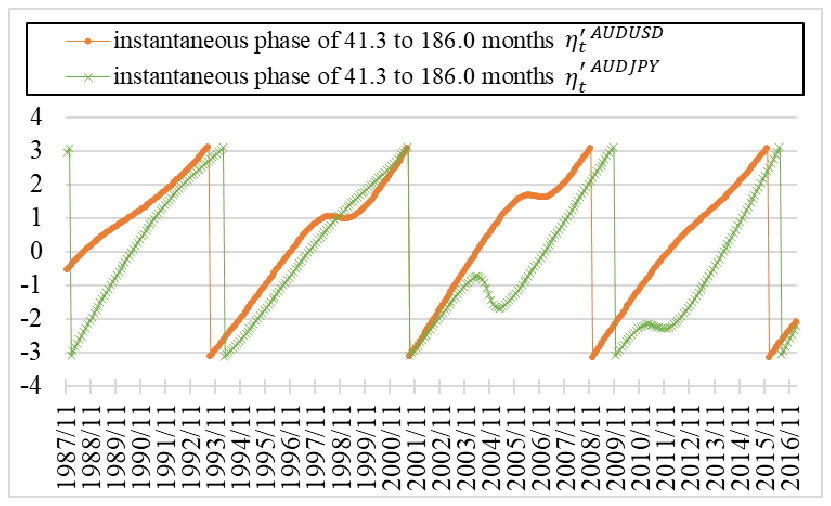}}
    \end{center}
    \begin{spacing}{1.1}
        Figure 7. (a) The instantaneous phase of 32.6--228.0 months $\eta^{\prime~AUDUSD}_t$ and $\eta^{\prime~AUDEUR}_t$. (b) The instantaneous phase of 38.0--228.0 months $\eta^{\prime~AUDUSD}_t$ and $\eta^{\prime~AUDEUR}_t$. (c) The instantaneous phase of 37.2--186.0 months $\eta^{\prime~AUDUSD}_t$ and $\eta^{\prime~AUDJPY}_t$. (d) The instantaneous phase of 41.3--186.0 months $\eta^{\prime~AUDUSD}_t$ and $\eta^{\prime~AUDJPY}_t$. Note: The phase differences in Figures 7(a)--(d) are not constant around (a) 2006, (b) 2006, (c) 2005 and 2012, and (d) 2005, respectively; however, all are almost constant during other periods.
    \end{spacing}
    \label{fig:f7}
\end{figure}

Figure 7 distinguishes between periods in which the phase difference is constant and fluctuates. Phase difference discontinuity affects its analysis. Therefore, an unwrapped instantaneous phase, which is defined, is used to allow for continuous change in the time development of an instantaneous phase.

The phase difference is expressed as
\begin{equation}
    \psi_t={\hat{\phi}}_t^{AUDUSD}-{\hat{\phi}}_t^{AUDj},
    \label{eq:Phase_d}
\end{equation}
where ${\hat{\phi}}_t^{AUDUSD}$ and ${\hat{\phi}}_t^{AUDj}$ denote unwrapped instantaneous phases of $\eta^{\prime~AUDUSD}_t$ and $\eta^{\prime~AUDj}_t$ at time $t$, respectively. We say that  $\eta^{\prime~AUDUSD}_t$ and $\eta^{\prime~AUDj}_t$ synchronize in the time interval $[t_0, t_1]$ if there exists a constant $d$ and a sufficiently small positive constant $\varepsilon$, such that:
\begin{equation}
    \left|\psi_t-d\right|<\varepsilon,
    \label{eq:syncro_e}
\end{equation}
for $t_0\le{{}^\forall}t\le t_1$.

\subsection{Synchronization index}
\label{Synchronization index}

The synchronization index $\gamma^2$ (\cite{Rosenblum_2001}) is employed for the time interval $1\le i\le W$ using the phase difference $\psi_i$ at time $i$ to measure the degree of synchronization between two time series,
\begin{equation}
    \gamma^2=\left(\frac{1}{W}\sum_{i=1}^{W}{\cos\psi_i}\right)^2+\left(\frac{1}{W}\sum_{i=1}^{W}{\sin\psi_i}\right)^2,
    \label{eq:Rosenblum1}
\end{equation}
Index $\gamma^2$ ranges from 0 to 1. When the phase difference between two time series is constant over time, $\psi_i$ takes a constant value. Thus, $\gamma^2$ takes the value close to 1 because $(\cos\psi_i,\mathrm{\ \sin}\psi_i)$ moves in the vicinity of one point on the unit circle. Therefore, if $\gamma^2$ is close to 1, then the two time series have high degrees of synchronization. In contrast, if $\gamma^2$ is close to 0, then the degree of synchronization is low.

Figure 8 shows the expected value of synchronization index $\gamma^2$ relevant to $W$. When $W = 13$ is selected, the expected value of $\gamma^2$ is 0.077 and the standard deviation of $\gamma^2$ is 0.074. Therefore, if the value of $\psi_i$ is given at random, then the synchronization index takes a value close to 0.077, which is a reference value for judging the randomness of $\psi_i$.

\begin{figure}
    \begin{center}
        \includegraphics[clip, width=0.42\columnwidth,height=0.28\columnwidth]{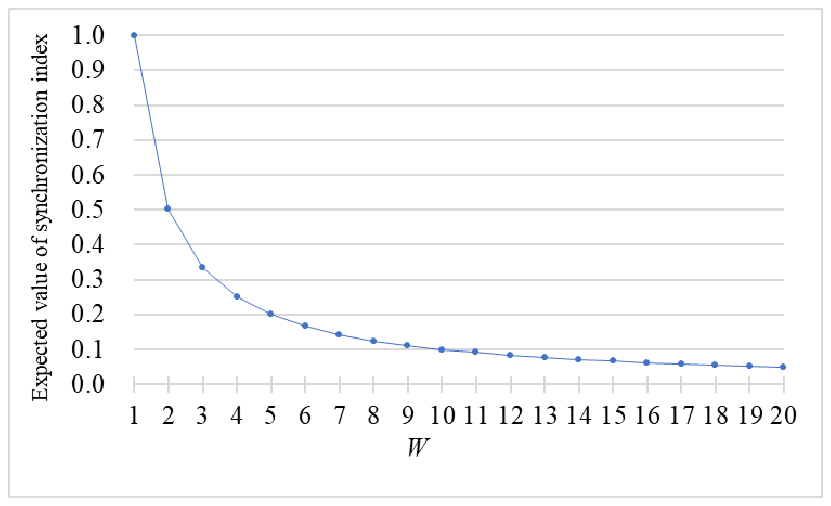}
    \end{center}
    \begin{spacing}{1.1}    
        Figure 8. The expected value of the synchronization index $\gamma^2$ with respect to $W$ calculated from 10,000 cases. $\psi_i$ is chosen from a uniform distribution of $[0,\ 2\pi]$. When we select $W = 13$, the expected value of $\gamma^2$ is 0.077 and the standard deviation of $\gamma^2$ is 0.074. The synchronization index with $W=13$ used in our analysis is small enough to include random values. Therefore, if the synchronization index is near 1, the two time series are considered to be highly synchronized.
    \end{spacing}
    \label{fig:f8}
\end{figure}

We measure the synchronization index for the time interval $W$ at each time $t$,
\begin{equation}
    \gamma_t^2=\left(\frac{1}{W}\sum_{i=t-p}^{t+p}{\cos\psi_i}\right)^2+\left(\frac{1}{W}\sum_{i=t-p}^{t+p}{\sin\psi_i}\right)^2,	
    \label{eq:Rosenblum2}
\end{equation}
where $p=(W-1)/2$ and $0<p<t$. In the following analysis, a window size of 13 (approximately one-year period) is set.
\footnote{
Changing the window size does not affect the result. The synchronization analysis is performed using four window sizes: 7 (approximately 0.5 year), 13 (approximately one year), 19 (approximately 1.5 years), and 25 (approximately two years). Although the amplitude magnitude is different, periods for which the synchronization index peaks do not change. In addition, periods for which the synchronization index maintains a high value do not change.
}

\section{Results and interpretation}
\label{Results and interpretation}

\subsection{Summary of the results}
\label{Result summary}

Figure 9 shows the synchronization index. If this index is stably high, the degree of synchronization between $\eta^{\prime~AUDUSD}_t$ (the AUD/USD exchange rate band-pass-filtered fluctuation around PPP, excluding the AUD inflation rate) and $\eta^{\prime~AUDj}_t$ is high. We call the synchronization index stably high when it maintains a high for a certain period, suggesting that the USD/currency j exchange rate fluctuates around the PPP level and confirming the establishment of PPP (see Figure 1). Conversely, if the synchronization index does not maintain a high level, the degree of synchronization between $\eta^{\prime~AUDUSD}_t$ and $\eta^{\prime~AUDj}_t$ is low. Although the synchronization index is temporarily high, by chance, the two instantaneous phase rhythms may match. Therefore, the degree of synchronization in such cases is not necessarily high. A low degree  of synchronization suggests that the USD/currency $j$ exchange rate fluctuated during this period because of factors other than PPP. For example, interest rate difference affect exchange rates. If the exchange rate fluctuates due to interest rate differences, then PPP may not hold. Therefore, the two time series have a low degree of synchronization in such periods.

Figure 9(a) shows the time development of the synchronization index  ${\hat{\gamma}}_p^2$ between $\eta^{\prime~AUDUSD}_t$ and  $\eta^{\prime~AUDEUR}_t$. The degree of synchronization is stably high during 2000:05--2005:08 and 2007:09--2013:04. This finding suggests that the USD/EUR exchange rate fluctuates around the PPP level in these periods. From 2002:04 to 2002:09, the degree of synchronization decreases slightly in the short term. Therefore, we consider that this period belongs to a stably high period. Similarly, from 2013:05 to 2014:07, the degree of synchronization decreases slightly. However, when we replace the numeraire, the degree of synchronization is significantly reduced (see Figure 10 in Appendix C). Therefore, this period is not stably high.

Figure 9(b) shows the time development of the synchronization index  $\hat{\gamma}_p^2$ between $\eta^{\prime~AUDUSD}_t$ and  $\eta^{\prime~AUDJPY}_t$. The degree of synchronization is stably high during 1991:08--1997:12, 1999:04--2003:10, and 2007:09--2010:10, suggesting that the USD/JPY exchange rate fluctuates around the PPP level during this period. From these results, the degree of synchronization  between $\eta^{\prime~AUDUSD}_t$ and $\eta^{\prime~AUDEUR}_t$ and between $\eta^{\prime~AUDUSD}_t$ and  $\eta^{\prime~AUDJPY}_t$ is, for the most part, high. Therefore, PPP holds in the frequency band used in this analysis. The degree of synchronization is low in certain periods for two reasons. First, an economic event may have an asymmetric effect on each $\eta^\prime_t$. Second, factors other than PPP affect exchange rates. For example, exchange rates change when the interest rate difference between two countries increases.

\begin{figure}
    \begin{center}
        \subfloat{({\bf a}) }{\includegraphics[clip, width=0.42\columnwidth,height=0.28\columnwidth]{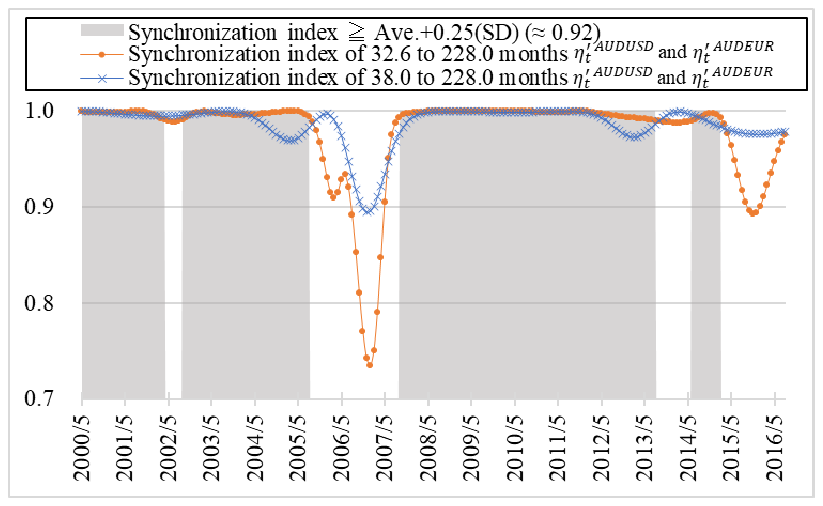}}
        \subfloat{({\bf b}) }{\includegraphics[clip, width=0.42\columnwidth,height=0.28\columnwidth]{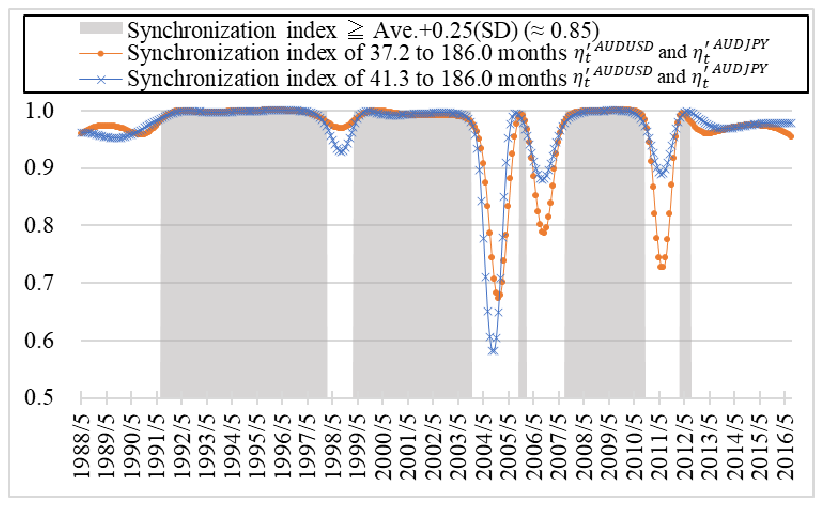}}\\
    \end{center}
    \begin{spacing}{1.1}
        Figure 9. (a) The synchronization index between $\eta^{\prime~AUDUSD}_t$ and $\eta^{\prime~AUDEUR}_t$. Note: The gray-colored interval indicates that the synchronization index from 32.6 to 228.0 months between $\eta^{\prime~AUDUSD}_t$ and $\eta^{\prime~AUDEUR}_t$ is the synchronization index $\gamma^2\geq\gamma^{2\ast}(\approx0.92)$, where $\gamma^{2\ast}$ denotes $\text{(the\ average\ value)}+0.25(\text{standard\ deviation})$. (b) The synchronization index from 37.2 to 186.0 months between $\eta^{\prime~AUDUSD}_t$ and $\eta^{\prime~AUDJPY}_t$. Note: The gray-colored interval indicates that the synchronization index between $\eta^{\prime~AUDUSD}_t$ and $\eta^{\prime~AUDJPY}_t$ is $\gamma^2\geq\gamma^{2\ast}(\approx0.85)$.
    \end{spacing}
    \label{fig:f9}
\end{figure}

\subsection{Interpretation of the results}
\label{Results interpretation}

This section discusses the relationship between fluctuations in the synchronization indices in the previous section and economic events. First, for the synchronization index between $\eta^{\prime~AUDUSD}_t$ and  $\eta^{\prime~AUDEUR}_t$, the synchronization index is stably high in many periods, suggesting the establishment of PPP. However, the degree of synchronization did not remain high during 2005:09--2007:08 and after 2013:05. During 2005--2006, a housing bubble and subsequent housing market collapse occurred in the United States, and housing prices frequently fluctuated during this period. Therefore, PPP cannot explain the price index fluctuation in the United States during this period. Because an asymmetric event occurred, the degree of synchronization did not maintain a high level during 2005:09--2007:08. Since approximately 2014, the EUR has depreciated against the USD. In addition, the exchange rate from 2015 to the first half of 2017 has been approximately 1 USD = 0.9 EUR and continues to diverge from the PPP level. The exchange rate moved toward the PPP level in the second half of 2017 (Figure 2(a)). The EUR has depreciated since 2014 because of interest rate reductions, the introduction of negative interest rates, and quantitative easing by the European Central Bank. PPP may not explain these exchange rate fluctuations. During the divergence period from the PPP level this exchange rate overlapped with the period during which the degree of synchronization did not maintain a high level after 2013:05. The Lehman Brothers’ bankruptcy in 2008:09 marked the beginning of a worldwide recession. During the economic crisis, countries’ economic variables tended to move in the same direction. Therefore, the degree of synchronization was high during 2008:9 and might not have been affected by PPP.

Second, the synchronization index between $\eta^{\prime~AUDUSD}_t$ and $\eta^{\prime~AUDJPY}_t$ is stably high in several periods, suggesting the establishment of PPP. However, the degree of synchronization does not remain high from 1988:05 to 1991:07, approximately during 1999, from 2003:11 to 2007:08, and after 2010:11. From 1986 to 1989, Japan experienced an economic bubble, and stocks and real estate prices soared. In addition, a technology bubble occurred around 1999. These events may have hindered the establishment of PPP from 1988:06 to 1991:07 and approximately during 1999. During 2005--2006, the United States experienced a housing bubble and subsequent housing market collapse. The event period overlaps with the period during which the degree of synchronization was not high between 2003:11 and 2007:08. The Great East Japan Earthquake in 2011:03 caused violent JPY fluctuations. During this period, the JPY appreciated because of the JPY purchase for insurance claim payments following the earthquake. In addition, investors may have bought JPY in anticipation of the currency strengthening. During 2011, the degree of synchronization was low, which may have resulted from JPY fluctuations. Moreover, since 2013, the JPY has depreciated against the USD, when the exchange rate remained deviated from the PPP level (see Figure 2(c)), because of the Bank of Japan’s quantitative easing and a 2011 rebound to the highest JPY value. PPP may not explain this exchange rate fluctuation. Therefore, the degree of synchronization did not remain high after 2010:11. The Lehman Brothers collapse in 2008:09 initiated a worldwide recession. Countries’ economic variables during the economic crisis tended to move in the same direction. Therefore, the degree of synchronization was high during 2008:09 and might not have been affected by PPP.

\section{Comparison with correlation coefficient}
\label{Comparison with the correlation coefficient}

A correlation coefficient is often used in economic studies to measure the strength of the relationship between the movements of two time series. However, the time difference in phase between two synchronized time series affects correlation coefficients. In contrast, the synchronization index is not affected by the time difference in phase. Figure 10(a) shows the synchronization index and correlation coefficient between $\sin(2\pi t)$ and $\sin(2\pi(t+\mathrm{\Delta}t))$ relevant to $\mathrm{\Delta}t$. The synchronization index is 1 regardless of the existence of the time difference in phase. Depending on the $\mathrm{\Delta}t$ size, the correlation coefficient can take any value between $-1$ and 1. The synchronization index is useful for measuring the synchronization of two time series with the time difference in phase.

\begin{figure}
    \begin{center}
        \subfloat{({\bf a})}{\includegraphics[clip, width=0.42\columnwidth,height=0.28\columnwidth]{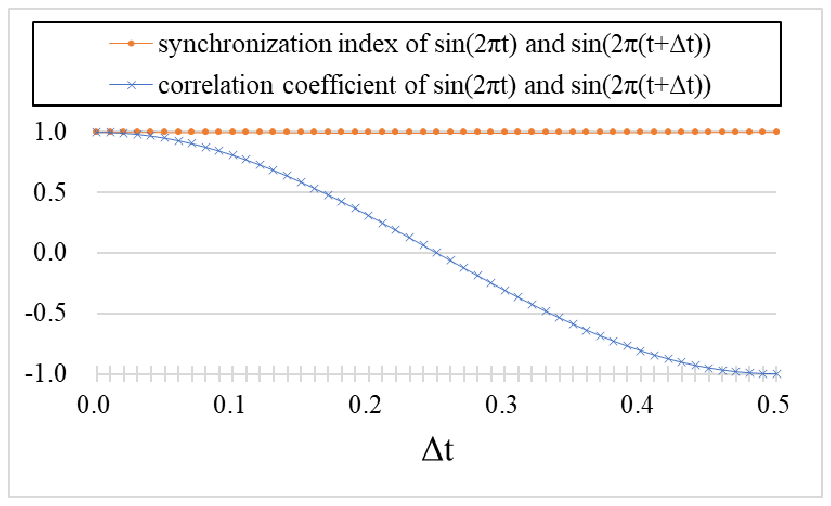}}\\
        \vspace{2mm}
        \subfloat{({\bf b}) }{\includegraphics[clip, width=0.95\columnwidth,height=0.28\columnwidth]{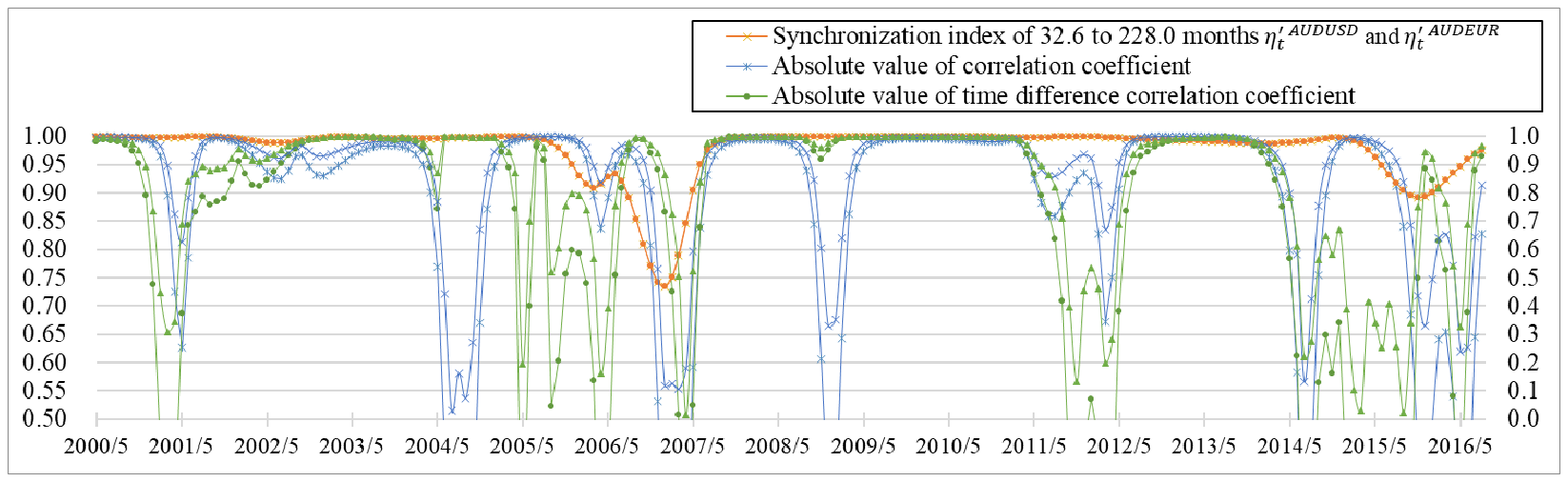}}\\
        \vspace{2mm}
        \subfloat{({\bf c}) }{\includegraphics[clip, width=0.95\columnwidth,height=0.28\columnwidth]{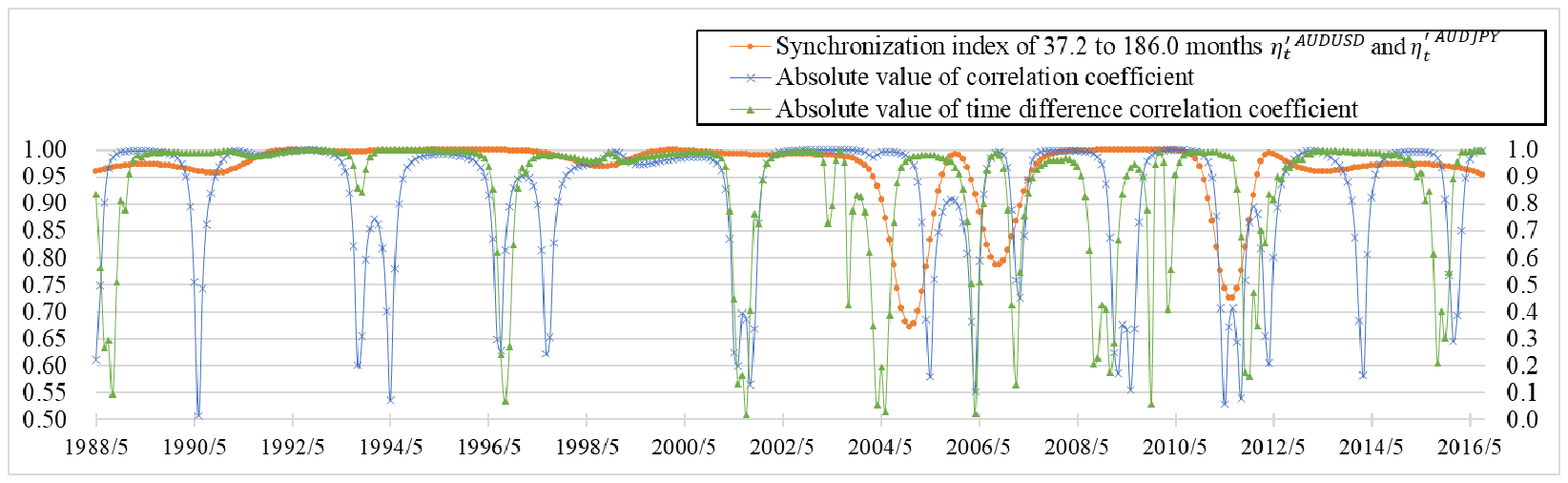}}
    \end{center}
    \begin{spacing}{1.1}    
        Figure 10. (a) The synchronization index and correlation coefficient between two time series $\sin(2\pi t)$ and $\sin(2\pi(t+\mathrm{\Delta t}))$. (b) The synchronization index and correlation coefficient from 32.6 to 228.0 months between $\eta^{\prime~AUDUSD}_t$ and $\eta^{\prime~AUDEUR}_t$. Note: “$\bullet$” (orange), “$\times$” (blue), and “$\blacktriangle$” (green) represent the synchronization index, absolute value of the correlation coefficient, and absolute value of the correlation coefficient  from 32.6 to 228.0 months with the time difference, respectively, between $\eta^{\prime~AUDUSD}_t$ and $\eta^{\prime~AUDEUR}_t$. (c) The synchronization index and correlation coefficient from 37.2 to 186.0 months between $\eta^{\prime~AUDUSD}_t$ and $\eta^{\prime~AUDJPY}_t$. Note: “$\bullet$” (orange), “$\times$” (blue), and “$\blacktriangle$” (green) represent the synchronization index, absolute value of the correlation coefficient, and absolute value of the correlation coefficient from 37.2 to 186.0 months with the time difference, respectively, between $\eta^{\prime~AUDUSD}_t$ and $\eta^{\prime~AUDJPY}_t$.
    \end{spacing}
    \label{fig:f10}
\end{figure}

If the time difference in phase of two time series does not significantly change over time or is clearly known inn each period, then the correlation coefficient can be used by shifting the time series with the time difference in phase. However, an appropriate time difference is difficult to determine from our data. We compare the synchronization index and correlation coefficient with the time difference in phase between $\eta^{\prime~AUDUSD}_t$ and $\eta^{\prime~AUDj}_t$. Figures 10(b) and (c) show the time series of the synchronization index and the absolute value of a correlation coefficient and that of a correlation coefficient with the time difference in phase.\footnote{
See Appendix D for the calculation method of the correlation coefficient with the time difference in phase.
}
The correlation coefficient is calculated using the same moving window (window size $W$ = 13) as that of the synchronization index. In addition, the time difference correlation coefficient considers the time difference in phase between two time series. Figure 10(b) shows the synchronization index and correlation coefficient between $\eta^{\prime~AUDUSD}_t$ and $\eta^{\prime~AUDEUR}_t$ from 32.6 to 228.0 months. Figure 10(c) shows the synchronization index and correlation coefficient between $\eta^{\prime~AUDUSD}_t$ and $\eta^{\prime~AUDJPY}_t$ from 37.2 to 186.0 months. Using these figures, when the synchronization index greatly decreases during a period, the correlation coefficient also decreases. The time series of a synchronization index and correlation coefficient sometimes behave differently when the synchronization index is stably high. An appropriate time difference in phase is difficult to identify from our data in which the lead-lag relationship varies during the short term. Thus, we employ the synchronization index.

In addition to the correlation coefficient, many other methods exist to analyze the relationship between two time series. For example, cross-correlation is often used. The differences between each of these methods and the synchronization index are as follows. Using the cross-correlation function:
\begin{equation}
    R_\tau^{xy} = \int_\infty^\infty x_t y_{t-\tau}~dt,
    \label{eq:cross-correlation}
\end{equation}
we can measure the similarity between two time series $x, y$ with time difference $\tau$. However, we consider that the economic data used in our analysis are likely to vary in structure with economic shocks and changes in economic conditions. Therefore, we can expect that the time lag of the relationship between two time series can change over time. Identifying the time lag in advance is difficult and is the same problem as that of the correlation coefficient with the previously explained time differences. The synchronization method has an advantage in this aspect.

\section{Concluding remarks}
\label{Conclusions}

We determine that the degree of synchronization between $\eta^{\prime~AUDUSD}_t$ (band-pass-filtered fluctuation around the PPP of the AUD/USD exchange rate, excluding the AUD inflation rate) and $\eta^{\prime~AUDEUR}_t$ and between   $\eta^{\prime~AUDUSD}_t$ and $\eta^{\prime~AUDJPY}_t$ are high for most times. This result suggests that PPP holds in the long term and that the PPP level is the long-term equilibrium value of the USD/EUR and USD/JPY exchange rates. However, a high degree of synchronization is not maintained for several periods. This property can be attributed to the occurrence of asymmetrical economic events and factors other than PPP that affect the exchange rate, such as during the U.S. real estate bubble and the Great East Japan Earthquake.

Correlation coefficients are inappropriate in this study because of the frequent time difference in phase between two time series. In contrast, we use the synchronization index in our analysis to measure the synchronization degree without identifying the time difference value in phase at each time. Therefore, synchronization analysis is suitable for our study.

We are also interested in the currency that causes synchronization in each period. However, because the synchronization index cannot measure this synchronization, methods such as Granger causality are needed, which will be discussed in future work.

\section*{Acknowledgements}
\label{Acknowledgements}

The authors are grateful to Prof.~Eiji Ogawa and Prof.~Masao Kumamoto for their insightful comments and helpful suggestions. 
They would like to thank the anonymous referees for helpful comments.
This work was partly supported by JST PRESTO (JPMJPR16E5), JSPS KAKENHI (17K05360, 19K01593, 19KK0067 and 21K18584), Tokio Marine Kagami Memorial Foundation, and Asset Management One Co., Ltd.

\section*{Appendix A. Theories of Exchange Rate Determination}
\label{Appendix A. Theories of Exchange Rate Determination}

Many factors affect exchange rates. Typical examples include price levels, interest rates, and balances of payment. Here, we describe exchange rate fluctuations on the basis of these factors.
As a result of international commodity arbitrage, the law of one price is established.
\begin{equation}
    S_t^{xy}P_t^x=P_t^y.
    \label{eq:APPP}
\end{equation}
where $S_t^{xy}$ denotes the $x/y$ exchange rate at time $t$, $P_t^x$ denotes the price level of the  country of currency $x$ at time $t$, and $P_t^y$ denotes the price level of the country of currency $y$ at time $t$.
The absolute PPP determines the exchange rate level from the ratio of the price levels between the two countries.
\begin{equation}
    S_t^{xy}=\frac{P_t^y}{P_t^x}.
    \label{eq:APPP}
\end{equation}
The relative PPP determines the rate of change in the exchange rate from the difference in inflation rates between the two countries.
\begin{equation}
    \frac{S_t^{xy}-S_{t-1}^{xy}}{S_{t-1}^{xy}}=\dot{P}_t^y-\dot{P}_t^x.
    \label{eq:RPPP}
\end{equation}
where $\dot{P}_t^x$ denotes inflation rate of the country of currency $x$ at time $t$, and $\dot{P}_t^y$ denotes the inflation rate of the country of currency $y$ at time $t$.

The uncovered IRP determines the rate of change in the exchange rate from the difference in interest rates between the two countries.
\begin{equation}
    \frac{S_t^{xy}-S_{t-1}^{xy}}{S_{t-1}^{xy}}=i_t^y-i_t^x.
    \label{eq:RPPP}
\end{equation}
where $i_t^x$ denotes the interest rate of the country of currency $x$ at time $t$, and $i_t^y$ denotes the interest rate of the country of currency $y$ at time $t$.

Imbalances in the balance of payments affect exchange rate fluctuations. For example, a current account surplus in the home country increases the demand for the home currency, causing the home currency to appreciate. Conversely, a home country's current account deficit increases the supply of its currency, causing the home currency to depreciate.

This paper focuses on exchange rate deviations on the basis of the relative PPP. The deviations are expected to be influenced by long-run adjustments in price levels and the productivity between the two countries. Therefore, our analysis focuses on the long-term movement of exchange rates on the basis of PPP.

Interest rates and balance of payments are also considered to affect the long-term movement of exchange rates. We follow two steps to focus mainly on the relationship between PPP and the exchange rate. First, in our analysis, we use data on exchange rate deviations from PPP. Second, we estimate the frequency band that corresponds to adjustments in the exchange rate to PPP and extract the frequency band from the data. The analysis including other factors, such as interest rates and balance of payments, is left for future work.

\section*{Appendix B. Fourier band-pass filter and power spectrum}
\label{Appendix B. Fourier band-pass filter and power spectrum}

We focus on recurrent patterns in a specific time scale to measure the degree of synchronization between two time series data. We employ a band-pass filter using a Fourier series representation and briefly review the Fourier series of function $f$. For simplicity, let $f$ be a real-valued continuous periodic function on $[0, L)$. The function $f$ can be represented as a Fourier series in Eq. (\ref{eq:A1}) as
\begin{equation}
    f(x)=\frac{a_0}{2}+\sum_{k=1}^{\infty}\left(a_k\cos\left(\frac{2\pi kx}{L}\right)+b_k\sin\left(\frac{2\pi kx}{L}\right)\right),	
    \label{eq:A1}
\end{equation}
where
\begin{equation}
    a_k=\frac{1}{L}\int_{0}^{L}{f(x)\cos\left(\frac{2\pi kx}{L}\right)dx\mathrm{\ }(k=0, 1, 2, 3, \cdots)},	
    \label{eq:A2}
\end{equation}
\begin{equation}
    b_k=\frac{1}{L}\int_{0}^{L}{f(x)\sin\left(\frac{2\pi kx}{L}\right)dx\mathrm{\ } (k=1, 2, 3, \cdots)}.	
    \label{eq:A3}
\end{equation}
We can consider the Fourier series for more general functions (e.g., Korner 2008). By taking a partial sum in Eq. (\ref{eq:A1}), we can create a band-pass-filtered periodic function $\widetilde{f}$ of a given function $f$ using bands $k$ for $1\le k_0\le k\le k_1$:  
\begin{equation}
    \widetilde{f}(x)=\sum_{k=k_0}^{k_1}{\left(a_k\cos\left(\frac{2\pi kx}{L}\right)+b_k\sin\left(\frac{2\pi kx}{L}\right)\right)\mathrm{\ } .}
    \label{eq:A4}
\end{equation}

The transformation procedure from discrete non-periodic time series data $g_n=g(x_0+n\mathrm{\Delta}x)\mathrm{\ }(n=0,\cdots, N-1)$ to band-pass-filtered discrete periodic time series data  ${\widetilde{f}}_n=\widetilde{f}(x_0+n\mathrm{\Delta}x)\mathrm{\ }(n=0,\cdots, N-1)$ is as follows.
\begin{enumerate}
    \item Using the linear transformation determined by $g_0$ at $x_0$ and $g_{N-1}$ at $x_{N-1}$, convert a given set of uniformly discretized $N+1$ time series data ${(g_n)}_{n=0,\cdots,N-1}$ into a periodic data  ${(f_n)}_{n=0,\cdots,N-1}$ such that $f_0=f_{N-1}(=g_0)$.
    \item Compute Fourier coefficients $a_k, b_k$ for ${(f_n)}_{n=0,\cdots,N-1}$.
    \item Construct a band-pass-filtered periodic time series data   ${({\widetilde{f}}_n)}_{n=0,\cdots,N-1}$ using $a_k, b_k$ for $k_0, k_1(1\le k_0\le k\le k_1)$.
\end{enumerate}
Step 1
\begin{equation}
    f_n=g_n-sx_n\mathrm{\Delta}x~~~~(n=0,...,N-1),
    \label{eq:A5}
\end{equation}
where $s={(g_{N-1}-g_0)}/{(x_{N-1}-x_0)},\mathrm{\ } x_n=x_0+n\mathrm{\Delta}x$ and $\mathrm{\Delta}x={(x_{N-1}-x_0)}/{(N-1)}$.
\\

\noindent Step 2

\noindent Compute
\begin{equation}
    a_k=\frac{1}{L}\sum_{n=0}^{N-1}{f_n\cos\left(\frac{2\pi k x_n}{L}\right)}\mathrm{\Delta}x~~~~(k=0,\cdots, K),
    \label{eq:A6}
\end{equation}
\begin{equation}
     b_k=\frac{1}{L}\sum_{n=0}^{N-1}{f_n\sin\left(\frac{2\pi k x_n}{L}\right)}\mathrm{\Delta}x~~~~(k=1,\cdots, K).
    \label{eq:A7}
\end{equation}
\\

\noindent Step 3
\begin{equation}
     {\widetilde{f}}_n=\sum_{k=k_0}^{k_1}{\left(a_k\cos\left(\frac{2\pi k x_n}{L}\right)+b_k\sin\left(\frac{2\pi k x_n}{L}\right)\right).}
    \label{eq:A8}
\end{equation}

\medskip

\noindent {\bf Power spectrum}: We can compute power spectrum $E(k)$ for each $k$,
\begin{equation}
     E(k)=\frac{a_k^2+b_k^2}{2}.
    \label{eq:A9}
\end{equation}
If $f$ is $C^l$ function, then $k^l\left|a_k\right|, k^l\left|b_k\right|<\infty$, implying that the Fourier coefficients decrease exponentially as $k$ increases. Notably, we plot $\sqrt{E(k)}$ and not $E(k)$ (Figure 3).

\section*{Appendix C. Robustness check of numeraire}
\label{Appendix C. Robustness check of numeraire}

The third country’s currency is introduced as a numeraire to analyze the synchronization between two currencies (e.g., USD and EUR). We use the AUD as the numeraire in the main body. We confirm the same results by using NZD as the numeraire in this appendix.

Figure 11(a) shows the time development of the synchronization index from 32.6 to 228.0 months between  $\eta^{\prime~AUDUSD}_t$ and $\eta^{\prime~AUDEUR}_t$ and between $\eta^{\prime~NZDUSD}_t$ and $\eta^{\prime~NZDEUR}_t$. The time series of the two synchronization indices behave similarly, except during 2014. Figure 11(b) shows the time development of the synchronization index from 37.2 to 186.0 months between $\eta^{\prime~AUDUSD}_t$ and $\eta^{\prime~AUDJPY}_t$ and between $\eta^{\prime~NZDUSD}_t$ and $\eta^{\prime~NZDJPY}_t$. The time series of the two synchronization indices behave similarly except during 1989:05.

\begin{figure}
    \begin{center}
        \subfloat{({\bf a}) }{\includegraphics[clip, width=0.42\columnwidth,height=0.28\columnwidth]{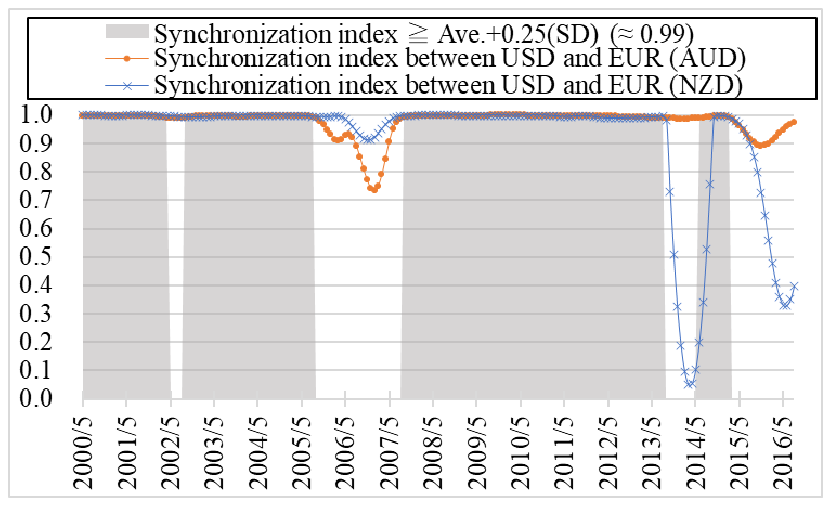}}
        \subfloat{({\bf b}) }{\includegraphics[clip, width=0.42\columnwidth,height=0.28\columnwidth]{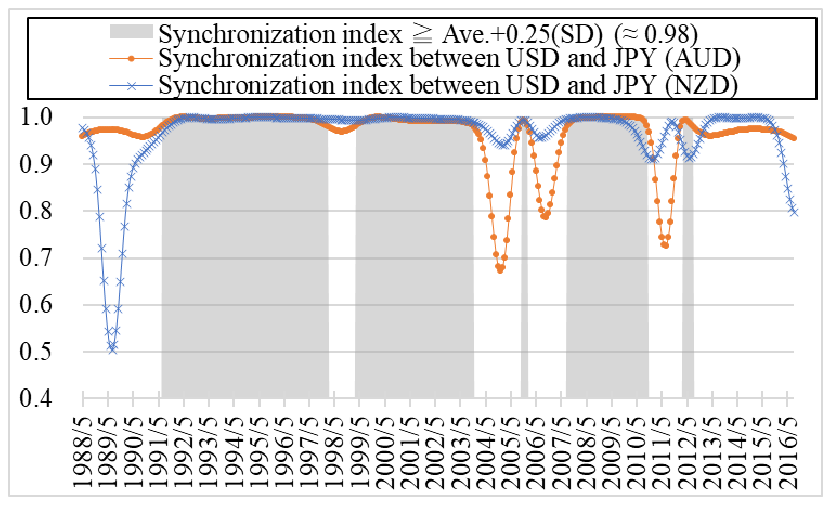}}\\
    \end{center}
    \begin{spacing}{1.1}
        Figure 11. (a) The synchronization index between $\eta^{\prime~AUDUSD}_t$ and $\eta^{\prime~AUDEUR}_t$ and between $\eta^{\prime~NZDUSD}_t$ and $\eta^{\prime~NZDEUR}_t$. Note: “$\bullet$” (orange) and “$\times$” (blue) represent the synchronization index from 32.6 to 228.0 months between $\eta^{\prime~AUDUSD}_t$ and $\eta^{\prime~AUDEUR}_t$ and between $\eta^{\prime~NZDUSD}_t$ and $\eta^{\prime~NZDEUR}_t$, respectively. The gray-colored interval indicates that the synchronization index from 32.6 to 228.0 months between $\eta^{\prime~AUDUSD}_t$ and $\eta^{\prime~AUDEUR}_t$ is the synchronization index $\gamma^2\geq\gamma^{2\ast}(\approx0.92)$, where $\gamma^{2\ast}$ denotes $\text{(the\ average\ value)}+0.25(\text{standard\ deviation})$. (b) Synchronization index between $\eta^{\prime~AUDUSD}_t$ and $\eta^{\prime~AUDJPY}_t$ and between $\eta^{\prime~NZDUSD}_t$ and  $\eta^{\prime~NZDJPY}_t$. Note: “$\bullet$” (orange) and “$\times$” (blue) represent the synchronization index from 37.2 to 186.0 months between $\eta^{\prime~AUDUSD}_t$ and $\eta^{\prime~AUDJPY}_t$ and between $\eta^{\prime~NZDUSD}_t$ and $\eta^{\prime~NZDJPY}_t$, respectively. The gray-colored interval indicates that the synchronization index from 37.2 to 186.0 months between $\eta^{\prime~AUDUSD}_t$ and $\eta^{\prime~AUDJPY}_t$ is $\gamma^2\geq\gamma^{2\ast}(\approx0.85)$.
    \end{spacing}
    \label{fig:f11}
\end{figure}

\section*{Appendix D. Calculation method for correlation coefficient with time difference in phase}
\label{Appendix D. Calculation method for a correlation coefficient with time difference in phase}

The time difference in phase between two time series affects a correlation coefficient. Thus, the correlation coefficient is not useful when the time difference is not sufficiently small. However, the correlation coefficient may show result similar to that of the synchronization index when the time difference in phase is considered. The correlation coefficient with the time difference in phase is calculated as follows.
\begin{enumerate}
    \item Identify the leading time series, and calculate the phase difference at each time. Find $m_t(={\hat{m}}_t)$ and $q_t(={\hat{q}}_t)$ that satisfy $\displaystyle \min_{q_t\in\mathbb{Z}}\left(\left|m_t\right|\right)$, where $m_t={\hat{\phi}}_t^{AUDUSD}-{\hat{\phi}}_t^{AUDj}+2q_t\pi$. The variables ${\hat{\phi}}_t^{AUDUSD}$ and ${\hat{\phi}}_t^{AUDj}$ denote unwrapped instantaneous phases of $\eta^{\prime~AUDUSD}_t$ and $\eta^{\prime~AUDj}_t$, respectively, both at time $t$.
    
    \item Calculate the time difference in phase at each time and ${\hat{l}}_t=l-1$ that satisfies
    \begin{equation}
        \begin{cases}
            \displaystyle \max_{l\in N, |l|\leq L}(\hat{\phi}_t^{AUDUSD}-\hat{\phi}_{t+l}^{AUDj}+2\hat{q}_t\pi)<0 & (\hat{m}_t>0)\\
            \displaystyle \min_{l\in N, |l|\leq L}(\hat{\phi}_{t+l}^{AUDUSD}-\hat{\phi}_t^{AUDj}+2\hat{q}_t\pi)>0 & (\hat{m}_t<0)
        \end{cases}
    ,
    \label{eq:C1}
    \end{equation}
    where $L$ is chosen depending on the data. The time difference in phase at each time is expressed as
    \begin{equation}
        \delta_t=
        \begin{cases}
            \hat{l}_t+\frac{\phi_t^{AUDUSD}-\phi_{t+\hat{l}_t}^{AUDj}}{\phi_{t+\hat{l}_t+1}^{AUDj}-\phi_{t+\hat{l}_t}^{AUDj}} & (\hat{m}_t>0)\\
            \hat{l}_t+\frac{\phi_t^{AUDj}-\phi_{t+\hat{l}_t}^{AUDUSD}}{\phi_{t+\hat{l}_t+1}^{AUDUSD}-\phi_{t+\hat{l}_t}^{AUDUSD}} & (\hat{m}_t<0)
        \end{cases}
    ,
    \label{eq:C2}
    \end{equation}
    
    \item Compute the absolute value of the correlation coefficient with the time difference in phase
    \begin{equation}
        \hat{r}_p^2=
        \begin{cases}
            \left|\frac{\displaystyle \displaystyle \sum_{i=t}^{t^\prime} \Bigl(\eta_i^{\prime AUDUSD}-\bar{\eta}^{\prime AUDUSD} \Bigr)\Bigl(\eta_{i+\bar{\delta}_t}^{\prime AUDj}-\bar{\eta}^{\prime AUDj}\Bigr)}{\sqrt{\displaystyle \sum_{i=t}^{t^\prime}\Bigl(\eta_i^{\prime AUDUSD}-\bar{\eta}^{\prime AUDUSD}\Bigr)^2}\sqrt{\displaystyle \sum_{i=t}^{t^\prime}\Bigl(\eta_{i+\bar{\delta}_t}^{\prime AUDj}-\bar{\eta}^{\prime AUDj}\Bigr)^2}}\right| & (\hat{m}_t>0)\\
            \\
            \left|\frac{\displaystyle \displaystyle \sum_{i=t}^{t^\prime} \Bigl(\eta_{i+\bar{\delta}_t}^{\prime AUDUSD}-\bar{\eta}^{\prime AUDUSD} \Bigr)\Bigl(\eta_i^{\prime AUDj}-\bar{\eta}^{\prime AUDj}\Bigr)}{\sqrt{\displaystyle \sum_{i=t}^{t^\prime}\Bigl(\eta_{i+\bar{\delta}_t}^{\prime AUDUSD}-\bar{\eta}^{\prime AUDUSD}\Bigr)^2}\sqrt{\displaystyle \sum_{i=t}^{t^\prime}\Bigl(\eta_i^{\prime AUDj}-\bar{\eta}^{\prime AUDj}\Bigr)^2}}\right| & (\hat{m}_t<0)
        \end{cases}
    .
    \label{eq:C3}
    \end{equation}

    where $t^\prime=t+W+1$, $p=t+\frac{W+1}{2}$ and ${\bar{\delta}}_t=\frac{1}{W}\sum\limits_{i=t}^{t+W-1}{\delta_i}$. $W$ denotes the window size.
\end{enumerate}

\end{document}